\documentclass{aastex}
\usepackage{graphicx}
\shorttitle{Accurate Centimeter-Wavelength Flux Density Scale}
\shortauthors{Perley and Butler}
\begin{document}
\title{An Accurate Flux Density Scale from 1 to 50 GHz}
\author{R. A. Perley and B. J. Butler}
\email{RPerley@nrao.edu, BButler@nrao.edu}
\affil{National Radio Astronomy Observatory}
\affil{P.O.Box O, Socorro, NM, 87801}
\slugcomment{Intended for the Astrophysical Journal, Supplement Series}

\begin{abstract}
We develop an absolute flux density scale for cm-wavelength astronomy
by combining accurate flux density ratios determined by the VLA
between the planet Mars and a set of potential calibrators with the
Rudy thermophysical emission model of Mars, adjusted to the absolute
scale established by WMAP.  The radio sources 3C123, 3C196, 3C286 and
3C295 are found to be varying at a level of less than $\sim$5\% per
century at all frequencies between 1 and 50 GHz, and hence are
suitable as flux density standards.  We present polynomial expressions
for their spectral flux densities, valid from 1 to 50 GHz, with
absolute accuracy estimated at 1 -- 3 \% depending on frequency. Of
the four sources, 3C286 is the most compact and has the flattest
spectral index, making it the most suitable object on which to
establish the spectral flux density scale.  The sources 3C48, 3C138,
3C147, NGC7027, NGC6542, and MWC349 show significant variability on
various timescales.  Polynomial coefficients for the spectral flux
density are developed for 3C48, 3C138, and 3C147 for each of the
seventeen observation dates, spanning 1983 through 2012.  The planets
Venus, Uranus, and Neptune are included in our observations, and we
derive their brightness temperatures over the same frequency range.
\end{abstract}


\maketitle

\section{Introduction}

The flux density scale most commonly used at centimeter wavelengths is
based on the spectral flux densities listed in \citet{Baa77} for the four
strong radio sources Cassiopeia A, Cygnus A, Taurus A, and Virgo A.
The data utilized in that paper were taken by many radio telescopes,
mostly in total power.  The measured spectral flux densities of these
four sources were termed by \citet{Baa77} as `absolute', meaning the
telescopes utilized had accurately known gains and the observations
were calibrated with a suitable thermodynamic standard. The
\citet{Baa77} paper provides convenient polynomial expressions for the
spectral flux densities for these four sources, valid between 30 MHz
and 30 GHz, with an accuracy estimated at about
5\%. However, because all four of these primary flux density sources
have angular sizes of several arcminutes, none is suitable as a
calibrator for modern high resolution interferometers.   

To provide a more widely distributed network of potential calibrators,
\citet{Baa77} also included polynomial expressions -- generally valid
from 400 MHz to 15 GHz, for 13 smaller diameter `secondary' sources
whose flux densities were determined by measuring their ratios to
Virgo A.  Of these 13 sources, three -- 3C48, 3C147, and 3C286, are
very compact, and have been extensively untilized by the VLA, and
other interferometers, for calibration.

Early observations taken by the VLA at 4885 and 1465 MHz suggested
that the ratios of the flux densities of 3C48, 3C147, and 3C286
deviated, at the level of a few percent, from ratios derived using the
\citet{Baa77} expressions.  As some, or all of these sources were
known to be variable, a program was started in 1983 to measure the
ratios of the flux densities of a set of small-diameter sources,
including six of the \citet{Baa77} secondary sources, at a range of
frequencies accessible to the VLA.  The goals were to correct, where
necessary, the \citet{Baa77} expressions for the adopted calibrators
(relative to one or more of them selected as the standard), and to
track any temporal variations of their flux densities.  As even the
smallest of these secondary sources are significantly resolved to the
VLA in its highest resolution configurations, most observations in
this program were conducted while in the most compact configuration,
with maximum spacing of $\sim$1 km.  This program -- modifed and
expanded as noted below -- has continued since 1983.

The majority of scientific observations taken in the 1980s were at the
6-cm and 20-cm receiver bands.  Beginning in the later 1980s, and
extending into the 1990s, significant improvements in VLA capabilities
were implemented as new receiver bands were added, three of the
initial four receiver bands were improved, and the technique of
referenced pointing was implemented, enabling much more precise high
frequency observations.  These changes required new calibration
objects to be added to this program, and a new flux density reference
source, valid for high frequencies, to be used, since the
\citet{Baa77} expressions are valid only to 15 GHz.

We will make frequent references in this paper to the VLA's receiver
bands, and to particular frequencies chosen for the observations.  The
appropriate nomenclature for these bands has long been a matter for
debate -- the common use of letter band codes, although convenient
from the engineering point of view, is not useful, and often confusing
for non-radio astronomers, especially since the same letters are often
used by other astronomical instruments for completely different
frequency bands.  In this paper, we will not utilize these letter
codes, except where their compactness is required in tables.  When we
are referring to a particular receiver system, we will refer to it as
a frequency band, and denominate it by the wavelength of the central
frequency of that band.  When we refer to a result which is specific
to a chosen frequency, we will denominate it by that frequency.  We
show in Table~\ref{tab:Nomenclature} the band names, the frequency
tuning ranges for each, and the commonly-used letter code.
\begin{deluxetable}{ccc}
\tabletypesize{\scriptsize}
\tablecolumns{3}
\tablecaption{VLA Frequency Band Nomenclature
\label{tab:Nomenclature}}
\tablewidth{0pt} 
\tablehead{
\colhead{Band Name}&\colhead{Band Code}&\colhead{Frequency Range (MHz)}}

\startdata 
90cm&P&300 -- 340\tablenotemark{a}\\ 
20cm&L&1000 -- 2000\\
10cm&S&2000 -- 4000\\ 
5cm &C&4000 -- 8000\\ 
3cm &X&8000 -- 12000\\ 
2cm&Ku&12000 -- 18000\\ 
1.3cm&K&18000 -- 26500\\ 
0.9cm&Ka&26500 --40000\\ 
0.7cm&Q&40000 -- 50000\\ 
\enddata 

\tablenotetext{a}{This is the tuning range of the old 90cm band system
which was disabled in 2009.  The VLA is currently being outfitted with
a new 90cm receiver system which spans 270 to 400 MHz.}
\end{deluxetable}

\section{Determining Flux Densities through Interferometry}

Single-dish total-power radio telescopes are conceptually simple, but
have distinct disadvantages for accurately measuring the flux
densities of radio sources, particularly at low frequencies.  Their
low resolution prevents discrimination against background objects or
extended emission located in the primary beam, so the observed power
includes unrelated sources whose contributions can only be estimated
statistically by observing nearby fields.  This problem is
particularly severe for sources embedded in a bright extended
background such as the galactic plane.  By contrast, an interferometer
functions as a spatial filter, discriminating against smooth
backgrounds, while at the same time, with its higher resolution,
enabling separation of the target source from the nearby confusing
sources.

The relation between the cross-product spectral power, $P_{corr}$,
provided by an interferometer, and the spectral flux density (or
visibility) $S$ of a source is expressed as (\citet{Per10}):
\begin{equation}
S = \frac{2k}{\eta_a\eta_e}\frac{1}{\sqrt{\epsilon_1A_{p1}\epsilon_2A_{p2}}}
\sqrt{\frac{T_{cal1}T_{cal2}}{P_{D1}P_{D2}}}P_{corr},
\end{equation}
where $\eta_a$ is the atmospheric attenuation, $\eta_e$ is the
efficiency of the digital electronics, including the correlator,
$A_{pn}$ and $\epsilon_n$ are the physical area and aperture
efficiency, respectively, of antenna $n$.  Calibration of the
electronics, including variations in the gain, is accomplished by
utilizing the value, $P_D$, measured at the correlator, of a signal
of power $P_{cal}=kT_{cal}\Delta\nu$ injected at the receiver.

Equation (1) could be used to directly determine the flux density of a
source provided all the factors were known with sufficient accuracy.
But in fact there are significant uncertainties in the determination
of some of these:
\begin{itemize}
\item Correction for system gains, and changes in the gains, requires
  accurate knowledge of the switched power, $P_{cal}$.  These values
  are measured in the lab, but after deployment in the field, the
  actual values may vary by a few percent, due to variations over time
  and temperature. Furthermore, because the switched power measurement
  is a true total power measurement, even low levels of
  radio-frequency interference (RFI) can perturb the measurements of
  $P_D$.  This is a significant problem for the lower frequency bands.
\item The antenna aperture efficiency, $\epsilon$, measures the
  fraction of the total incident source power intercepted by an
  antenna which is delivered to the receiver.  Factors which reduce
  efficiency include blockage and refraction of the incoming signal by
  the support legs and subreflector, the illumination taper introduced
  by the feed, mispointing of the antenna, errors in optical
  alignment, and distortions of the primary reflector. In general, the
  efficiency of the VLA's antennas is known with an accuracy of
  $\sim$5\%.  The antenna efficiency is a strong function of elevation
  at the higher frequencies due to gravitationally-induced distortions
  of the antenna surface and quadrupod support legs.
\item The atmospheric absorption term, $\eta_a$, has a strong
  elevation dependency at higher frequencies, and can be time and
  spatially variable due to clouds.
\end{itemize}

Because of these uncertainties, direct use of eqn.(1) to convert the
interferometer products to flux density has not been practical for the
VLA if accuracies better than $\sim$10\% are desired.  However,
provided the system calibration powers are in error by only a constant
scale factor, and that the effects of antenna efficiency and
atmospheric absorption can be parameterized as time-independent
functions of elevation, the VLA can be used to determine highly
accurate flux density ratios between calibrators.  In the following
sections, we demonstrate how ratios with accuracies much better than
1\% can be determined at most frequencies, and how these ratios can be
used to determine accurate absolute flux densities by using the planet
Mars as the reference.

\section{The Methodology of Correcting Variable Gains}

The major factors which degrade the VLA's ability to accurately
measure flux density ratios are changing system gains, errors in
antenna pointing, and elevation dependencies of the antenna gain.  In
this section, we describe how each can be mitigated through careful
observing strategies and data post-processing.

\subsection{Accounting for Changing System Gains}

\begin{figure}[ht]
\centerline{\hbox{
\includegraphics[width=3in]{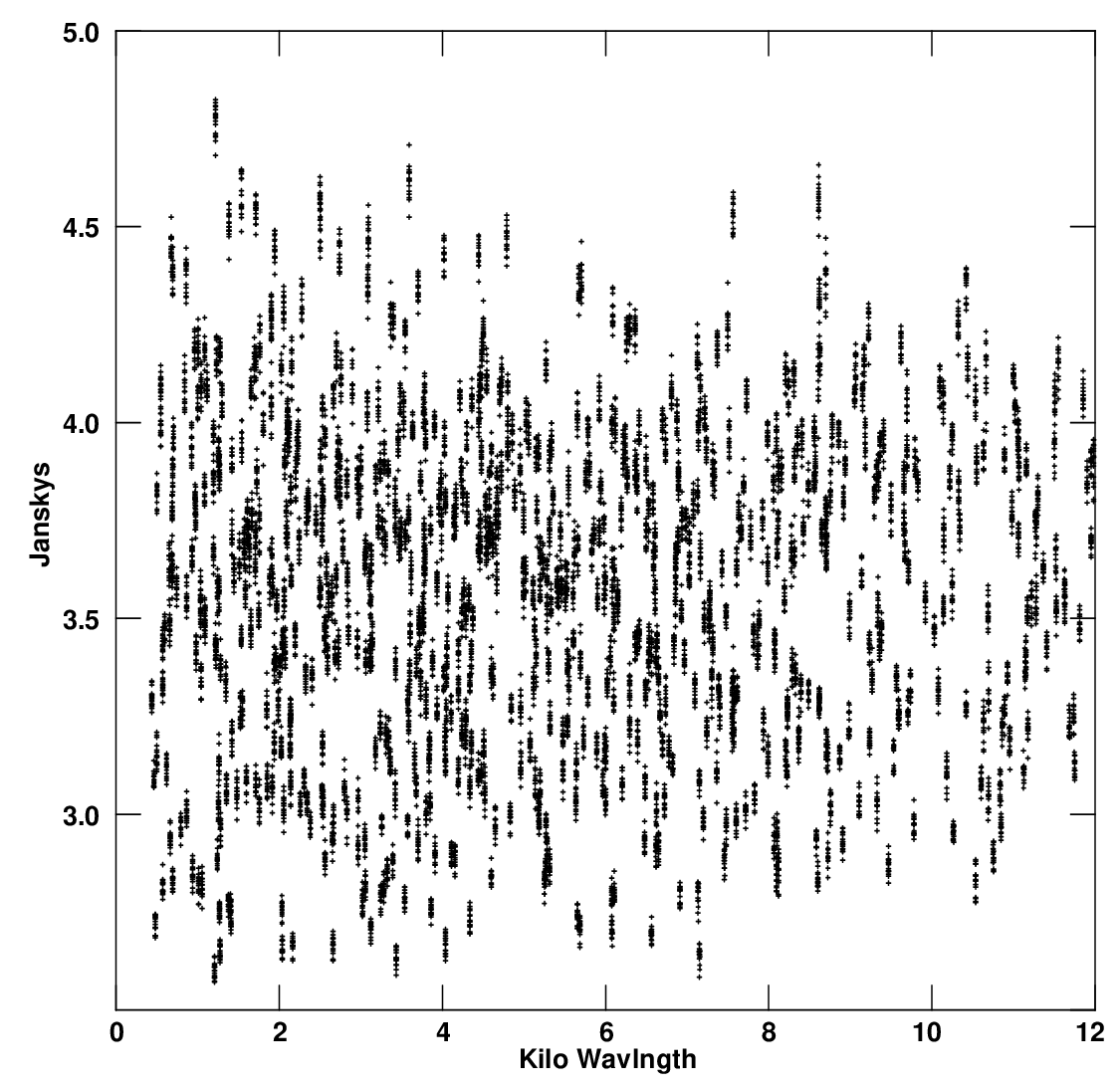}
\includegraphics[width=3in]{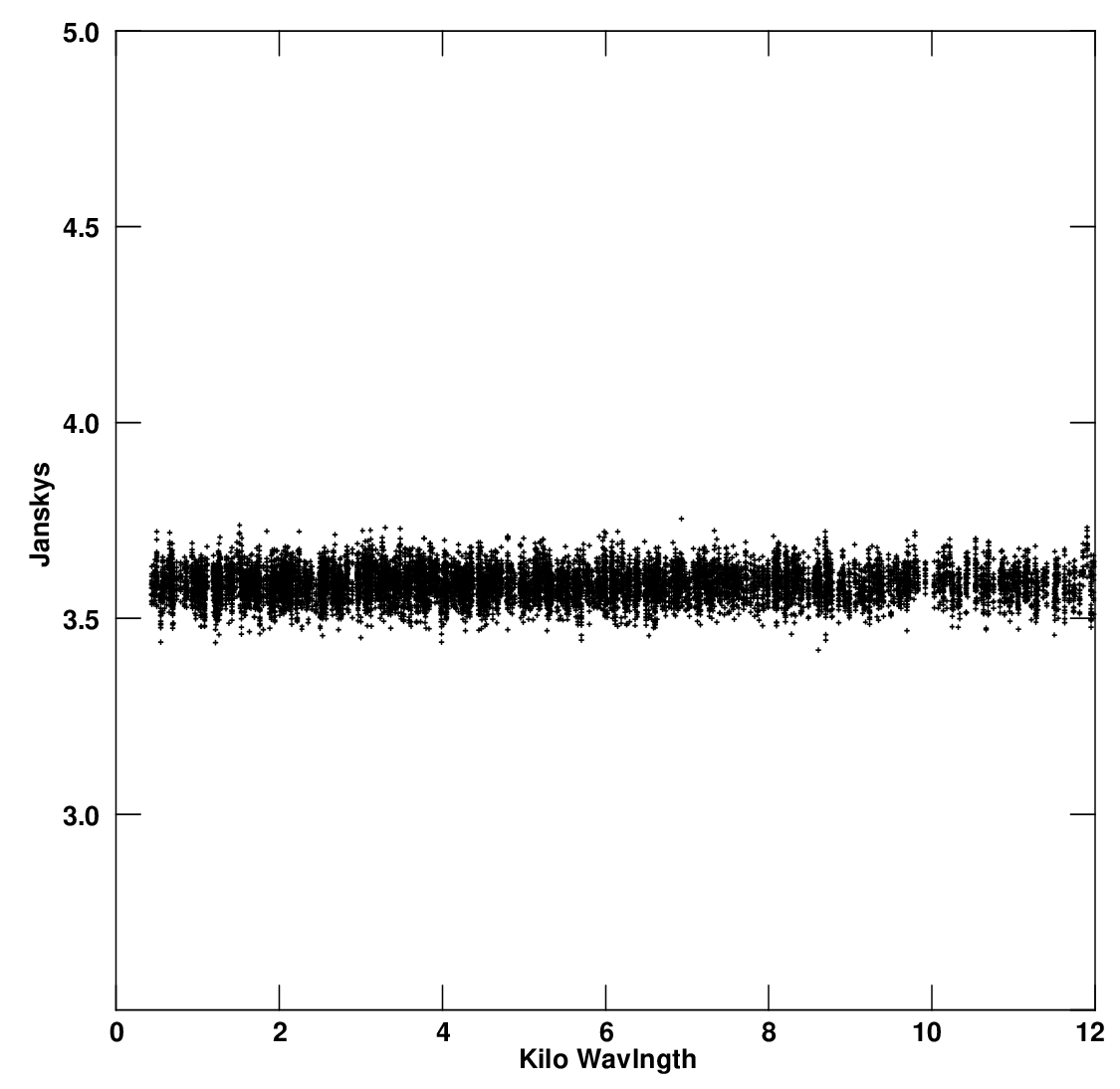}}}
\centerline{\parbox{6in}{
\caption{\small The left panel shows the visibilities for the
  calibrator J0217+7349 without correcting for system gain variations.
  The right panel shows the visibilities following correction for
  system gain changes using the switched power monitoring. Both plots
  are on the same scale to demonstrate the effectiveness of the
  switched power corrections.}
\label{fig:SysLin}}}
\end{figure}

To monitor system gains, a small, known amount of noise power is added
to the receivers, square-wave modulated at a frequency of 10 Hz.  This
additional power is synchronously detected and measured at the
correlator.  Changes in the detected calibration power reflect changes
in electronic gains, provided the injected power is itself stable.  To
demonstrate the effectiveness of this system, we show in
Figure~\ref{fig:SysLin} the correlation amplitudes (visibilities) for
a point source when the gain corrections have, and have not been
applied. These data were taken in clear weather over a 40-hour period
at a frequency of 4885 MHz, at which the elevation-dependent antenna
gain corrections and atmospheric opacity effects are negligible.
The application of the switched power correction has reduced the
apparent flux density variations from $\sim$50\% to less than 1\%.

\subsection{Correcting for Antenna Pointing}

Observations made in the early years of this program clearly indicated
that the primary source of error in the gain calibration process at
frequencies above $\sim$10 GHz is due to offsets in antenna pointing.
The accuracy of blind VLA pointing at night under calm conditions is
about 10 arcseconds.  On sunny days, pointing offsets induced by
differential solar heating of the antenna structure can exceed 1
arminute -- larger than the FWHM of the VLA primary beam at the
highest frequency band.

To remove such errors, the method of `referenced pointing' was
implemented on the VLA in the mid-1990s (\citet{Kes94}).  In this
method, a pointing determination is made on a nearby calibrator (or on
the target source itself, if sufficiently compact), and the offset
correction applied to subsequent observations.  Utilization of this
technique can reduce pointing error to as low as 3 arcseconds,
providing the calibrator object is close enough -- ideally within 5
degrees in azimuth and elevation, and preferably less than 15 degrees
-- and the weather calm and clear.  The offset determination is
normally made at a frequency within the 3cm band, as this has the best
combination of sensitivity and resolution, and the results applied to
observations made at other frequency bands.  This methodology was
first employed in our program in 1995, and was utilized for all
following observations.

\subsection {Correction for Elevation Gain Dependency}

\begin{figure}[ht]
\centerline{\hbox{
\includegraphics[width=3in]{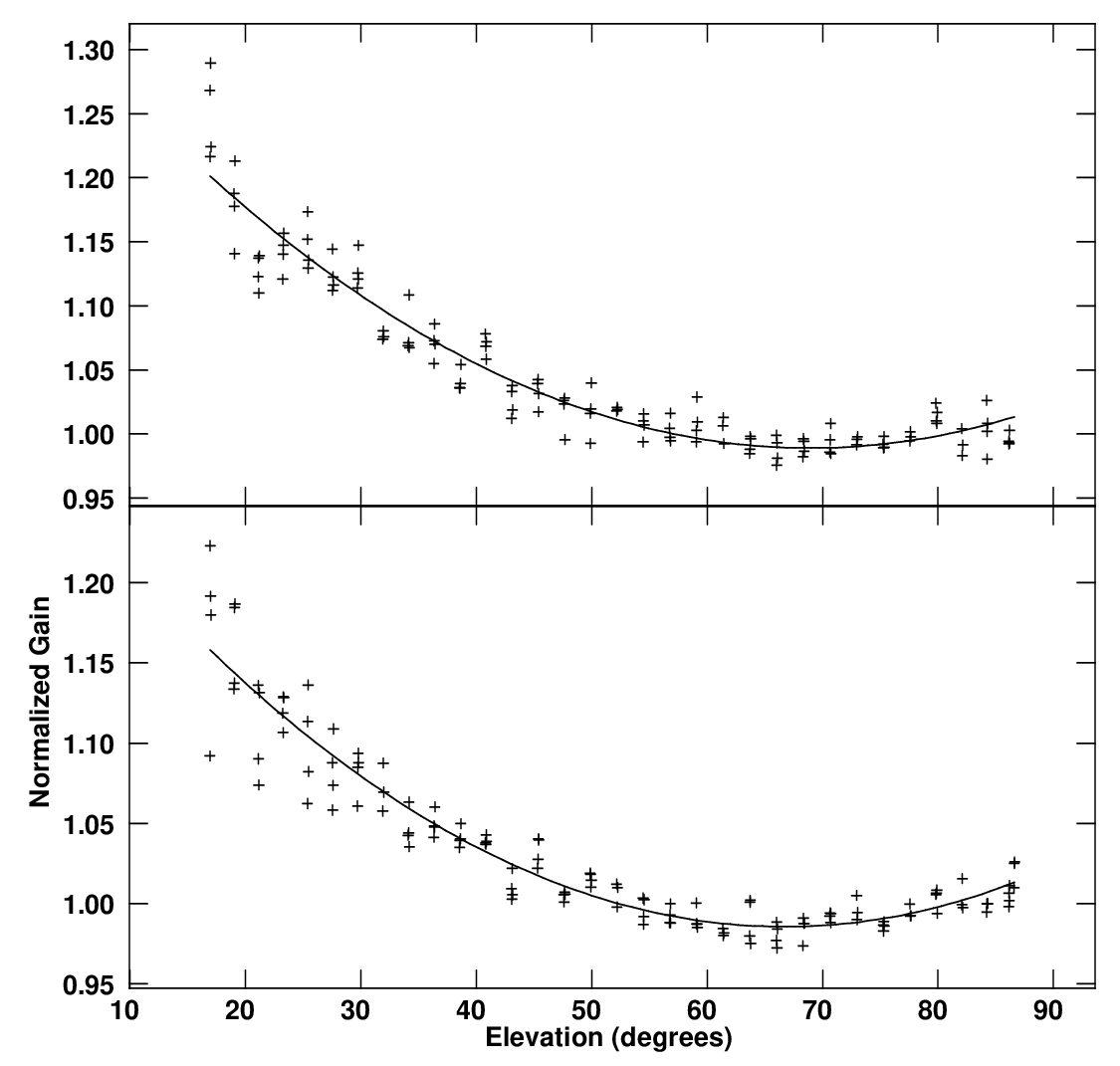}
\includegraphics[width=3in]{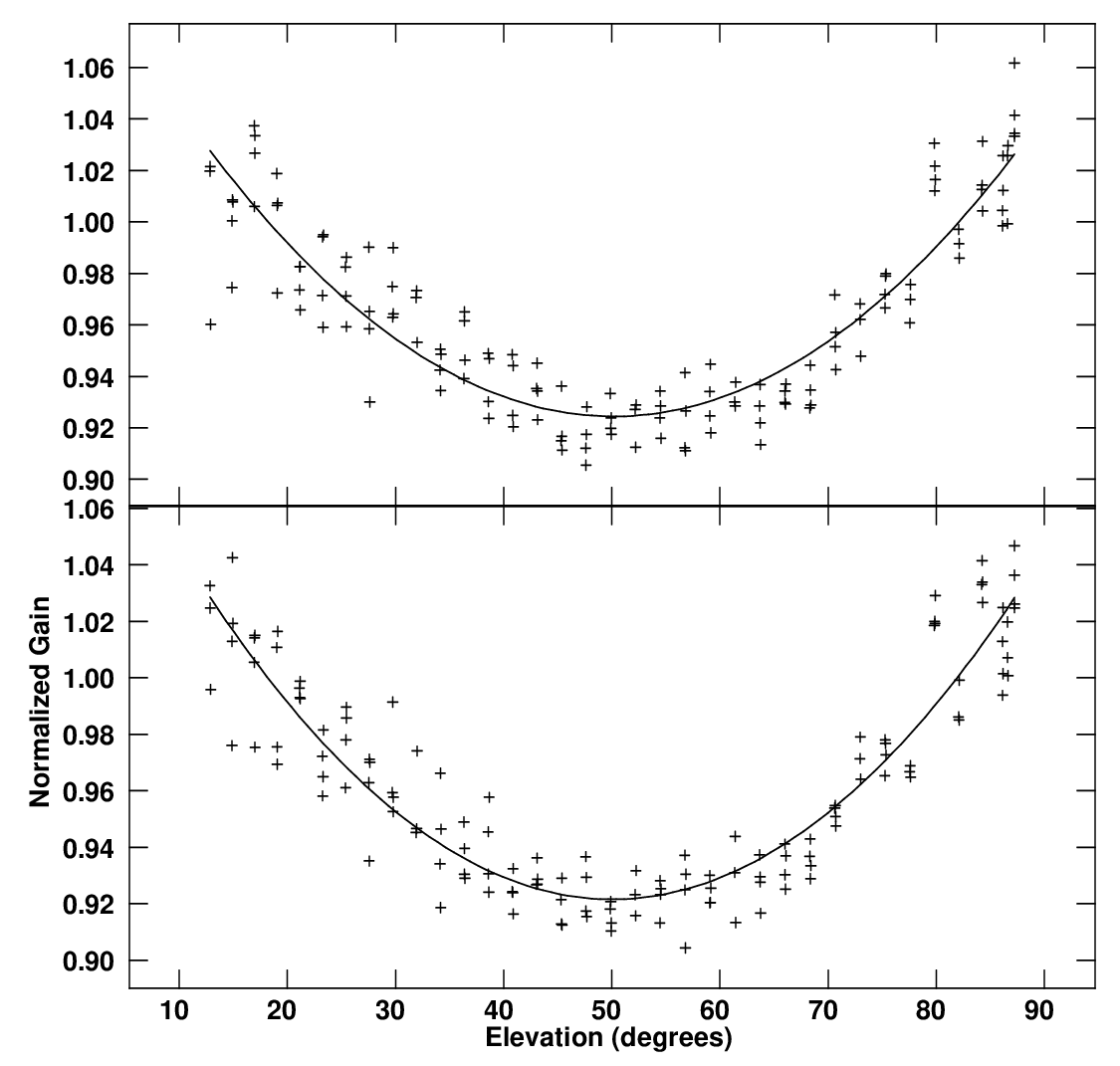}}}
\centerline{\parbox{6in}{
\caption{\small Examples of the elevation voltage gain dependencies at
  43 GHz.  The left panel pair shows the fit for both polarizations
  for antenna 21, the right panel pair for antenna 23. The post-fit
  rms power gain residuals are about 3\% for both antennas,
  corresponding to a pointing residual of four arcseconds.}
\label{fig:ElFit}}}
\end{figure}

The VLA antennas lose forward gain at high frequencies from a
combination of deformation of the antenna surface and a bending of the
quadrupod feed leg structure.  The loss of gain at high frequencies is
significant, but is repeatable and can be corrected for with good
accuracy, as illustrated by the fits shown in Figure~\ref{fig:ElFit},
taken from 43 GHz observations in October, 2011.
Referenced pointing was applied to these observations -- the scatter
about the best-fit gain represents the current limit for VLA pointing
-- approximately four arcseconds, rms.

\section{Source Selection}

The source list for the inaugural June 1983 observations comprised the
entire \citet{Baa77} source list, plus the compact source 3C138 which,
although known to be variable, is useful for calibration because of
its small angular size and high fractional linear polarization.  From
these, the seven most compact objects were selected for further
observations.  These were: 3C48, 3C123, 3C138, 3C147, 3C286, 3C295, and
NGC7027.  These sources were observed in all of the 19 separate
observation sessions.  
  
Over time, additional objects were added to the list: 3C196 in 1989,
NGC6572 in 2000, and MWC349 in 2001.  Beginning in 1995, observations
of the planets Mars, Venus, Jupiter, Uranus, and Neptune were added to
enable extension of the scale to the highest frequencies utilized by
the VLA, since the planets are commonly used as millimeter-wavelength
flux density standards (e.g. \citet{Gri93}).  Jupiter was dropped from
the list after four sessions, as it is too large for its total flux
density to be accurately measured by the VLA at the high frequency
bands.

We selected the planet Mars as the primary high-frequency standard.
Mars is an excellent source for this role, as it has a good emission
model (\citet{Rud87}, \citet{Muh91}, \citet{Sid00}), is usually free
of clouds and dust that can affect cm-wavelength observations, and is
usually small enough to be only partially resolved to the VLA in its
most compact configuration at all frequency bands.  The recent
publication of the absolute WMAP observations of Mars by \citet{Wei11}
has provided the capability of placing the entire cm-wavelength flux
density scale on an absolute standard.

Not all sources were observed at all frequency bands at all sessions,
as the planets are too weak to have been detected at frequencies below
$\sim$ 5 GHz prior to the VLA's upgrade, and 3C123 is too heavily
resolved at the highest frequencies.  As the VLA's sensitivity
improved dramatically following the implementation of the WIDAR
correlator in March 2010, subsequent observations included Venus,
Mars, Uranus and Neptune at all bands except at 90cm.

\section{Observations}

The observations were made in 19 observing sessions on the dates
listed in Table~\ref{tab:Log}.  The table also lists the time spent
for each session, the configuration in which the observations were
made, a summary of the available frequency bands, and a brief note
about the weather conditions.

\begin{deluxetable}{cccccccc}
\tabletypesize{\scriptsize} 
\tablecaption{Observing Log 
\label{tab:Log}}
\tablewidth{0pt} 
\tablehead{ \colhead{Date}& \colhead{IAT Start}&
\colhead{LST Start}& \colhead{Duration}& \colhead{Total}&
\colhead{Configuration}& \colhead{Bands}& \colhead{Comments}}
\startdata 

02-03 Jun 1983 &22:30&07:50&25&25&D&LCKuK&Clear\\ 
28-29 Dec 1985 &02:50&02:00&22&22&D&PLCKuK&Clear\\ 
04-05 May 1987 &07:45&15:20&24&24&D&PLCXKuK&Clear, calm\\ 
29-30 Dec 1989 &09:40&09:00&23&23&D&PLCXKuK&Cloudy\\ 
09 Jul 1992    &07:35&19:39&13&13&D&PLCXKuK&Cloudy\\ 
26 Aug 1992    &01:25&16:30& 5&5&D&PLCXKuK&T-storms\\ 
13-14 Mar 1995 &21:50&02:00&24&24&D&PLCXKuKQ&Partly cloudy\\ 
03-04 Feb 1998 &12:20&14:00&28&28&D&PLCXKuKQ&Cloudy, light snow\\ 
15-16 Apr 1999 &23:10&05:35&24&24&D&PLCXKuKQ&Clear\\ 
02-03 Oct 2000 &22:55&16:30&24&24&D&PLCXKuKQ&Clear\\ 
09-10 Nov 2001 &23:55&20:10&25&25&D&PLCXKuKQ&Partly cloudy\\ 
06-07 Feb 2003 &19:40&21:35&30&30&D&PLCXKuKQ&Mostly clear\\ 
27-30 Aug 2004 &01:30&16:35&27&34&D&PLCXKuKQ&Mostly clear\\ 
15-20 Jan 2006 &23:00&23:30&29&45&D&PLCXKuKQ&Windy\\ 
13-14 May 2007 &13:00&22:10&31&31&D&PLCXKuKQ&Clear\\ 
11-15 Sep 2008 &01:30&18:00&44&106&D&PLCXKuKQ&Mostly clear\\ 
09-11 Jan 2010 &08:50&09:00&31&36&D&LSCXKKaQ&Clear\\ 
26-27 Dec 2010 &16:30&15:45&30&30&C&LSCXKKuKKaQ&Clear\\ 
19-20 Jan 2012 &00:30&01:10&30&31&DnC&LSCXKKuKKaQ&Clear, Breezy\\ 

\enddata \tablecomments{The durations column gives the length of the
scheduled observing run in hours.  For some sessions, additional
observing time was obtained within a few days of the primary block, as
reflected by the difference between the Durations and Total
columns. Band names are coded for brevity: P=90cm, L=20cm, S=10cm,
C=6cm, X=3cm, Ku=2cm, K=1.3cm, Ka=0.9cm, Q=0.7cm. The longest baseline
in D configuration is $\sim$1 km, and $\sim$3 km in C configuration.}
\end{deluxetable}

\subsection{Frequency Selection}

The major goals of this program included both an accurate
determination of the variability of the sources, and an accurate
determination of the flux density ratios between the sources over as
wide a frequency range as possible.  As both the VLA's original, and
the new WIDAR correlator provide correlation products for two
different frequency tunings within each frequency band, we strove to
fix one of these tunings to a value which remained unchanged
throughout the duration of the project, and to place the other at the
maximum separation allowed by the electronics at the time of
observing.  Due to changes in the VLA's electronics over thirty years,
evolution of the goals of the project, and changes in our
understanding of the location and impact of RFI, the specific
frequencies utilized changed over the observation period, most notably
at the lower frequencies.  The frequencies chosen for all sessions up
to 2006 are given in Table~\ref{tab:FreqCoverage}.
\begin{deluxetable}{ccccccccccccccc}
\tabletypesize{\scriptsize}\tablecolumns{15}
\tablecaption{Detailed Frequency Settings, in MHz, from 1983 through 2006
\label{tab:FreqCoverage}}
\tablewidth{0pt} 
\tablehead{
\colhead{Year}&\colhead{P1}&\colhead{P2}&\colhead{L1}&
\colhead{L2}&\colhead{C1}&\colhead{C2}&\colhead{X1}&\colhead{X2}&
\colhead{U1}&\colhead{U2}&\colhead{K1}&\colhead{K2}&\colhead{Q1}&
\colhead{Q2}}

\startdata 
1983   &   &   &    &1465&    &4885&    &    &     &14965&     &22485&&\\
1985   &317&333&1425&1465&4835&4885&    &    &14915&14965&22435&22485&&\\
1987   &327&333&1425&1465&4816&4866&8435&8485&14934&14984&22435&22485&&\\
1989   &327&333&1425&1465&4816&4866&8435&8485&14934&14984&22435&22485&&\\
1992   &327&333&1425&1465&4816&4866&8435&8485&14934&14984&22435&22485&&\\
1995   &327&333&1365&1475&4835&4885&8415&8465&14915&14965&22435&22485&43315&43365\\
1998   &321&327&1365&1475&4835&4885&8435&8485&14915&14965&22435&22485&43315&43365\\
1999   &321&327&1275&1465&4835&4885&8435&8485&14915&14965&22435&22485&43315&43365\\
2000-2006&321&327&1275&1465&4535&4885&8435&8735&14915&14965&22435&22485&43315&43365\\
\enddata

\tablecomments{Additional frequencies were added in the 2007 and 2008
sessions to allow more uniform spectral coverage.  Frequency
selections for the wideband WIDAR correlator (sessions 2011 and 2012)
were set to include the frequencies shown in the last row of the
table.  No 90cm band observations were made after the 2008 session, nor
in the 2cm band in 2010, due to EVLA construction.}
\end{deluxetable}
Major changes in VLA electronics due to the VLA's upgrade permitted a
much wider range in frequency selection for sessions after 2006.
Observations after this date added new frequencies, but always
retained those listed in the last line of
Table~\ref{tab:FreqCoverage}.  The 90cm band observing system was
disabled in 2009, so no observations were possible at this band
following the 2008 session.  Due to the implementation of new
receivers, no 2 cm band data were taken in the 2010 session.

\subsection{Observing Methodology}

The general methodology was to observe each source approximately
hourly at each band.  To prevent holes in coverage, the durations of
most sessions were at least 24 hours.  Given the restrictions in
antenna slew rate, source separation, and source elevation, we
typically obtained five to ten individual `snapshot' observations of
each object at each band, with each observation's duration being 30 to
60 seconds.  This methodology permits accurate estimation of and
correction for elevation-dependent telescope characteristics, and
allows a good estimation of the errors in the flux density.  This
multiple observation strategy also ensures minimal impact on the
project goals in case of short-duration telescope malfunctions, bad
weather, and sporadic interference.  To permit careful editing of
discrepant data, the visibility integration time was kept short,
typically 3.3 seconds or less.

\subsection{Calibration}

All editing, calibration, imaging, and image analysis were done within
the {\tt AIPS} software package (\citet{Gre04}).  The same editing and
calibration methodology was employed for every source at each band for
every session, to maximize uniformity and reliability of the results.
The sequence of operations was as follows:
\begin{enumerate}
\item {\bf Initial editing:} Most sources are strong enough to allow a
  simple initial editing to purge obviously invalid data -- typically
  showing up as abnormally low amplitudes.  On occasion, RFI or solar
  interference was seen, with the affected data removed in the same
  way.
\item {\bf Estimation of atmospheric opacity:} For the observations
  made between 1995 and 2007, antenna dips were done every six hours
  at 1.3cm and 0.7cm bands to permit estimation of atmospheric opacity
  (\citet{But96}).  These values, or the values from an atmospheric
  model using surface and seasonal weather data for the years where no
  dip data were available (\citet{But02}), were applied to the
  visibility data.
\item {\bf Removal of atmospheric and instrumental phase
  perturbations:} Phase fluctuations were removed by application of
  phase self-calibration for each source, using a point-source model
  for the unresolved sources, or a `clean-component' model for
  resolved sources.  In the latter case, the model was derived from
  the data, using an imaging/self-calibration loop.  For both, the
  phase solutions were determined for each time integration, and the
  results applied to the corresponding visibility data.  The SNR for
  the low-frequency planetary observations was generally too low to
  permit this procedure -- however, at these frequencies, the phase
  stability was always good enough that such detailed phase
  fluctuation removal was not necessary.
\item {\bf Removal of data taken with bad pointing:} The most
  difficult error to detect -- and the error which ultimately limits
  the accuracy of high frequency observing -- was that due to residual
  pointing errors.  We did this by first making an amplitude
  calibration solution for each source, using either a point-source
  model, or a `clean-component' model.  We then examined these
  solutions, flagging those data for which the corresponding power
  gains deviated by more than a fixed fraction, varying from 1\% for
  the low frequency bands to 25\% for the highest frequency band.  In
  nearly all cases, the flagging is on data whose amplitudes are too
  low -- consistent with the cause being pointing offsets. To prevent
  bias, all sources at any one band were flagged with the same
  threshold.  We emphasize that these amplitude gain solutions were
  never applied to the data -- they were utilized only to identify bad
  data.
  \item {\bf Estimation and removal of elevation gain dependency:} The
  elevation gain dependency was estimated by jointly fitting the
  calibration gain solutions for the source flux densities and an
  antenna-based elevation dependency of the form: $G = G_0 +
  G_1\sec(z) + G_2\sec^2(z)$. The resulting gain function was then
  applied to all the visibility data.  For all but the last two
  sessions, the objects utilized were the four sources which are
  nearly unresolved at all frequencies in our observations -- 3C48,
  3C138, 3C147, and 3C286.  For the final two sessions, approximately
  30 additional point-sources were added to the source list and were
  included in this calibration step.  Separation of the elevation gain
  dependency from individual pointing offsets required two or three
  interations of this and the previous step.
\item {\bf Final gain calibration:} After all discrepant data were
  removed, and the elevation dependency measured and corrected for, a
  final amplitude calibration solution was made using a single source,
  either 3C147 or 3C286.  The average gain from these solutions was
  then applied, without any temporal trend, to all sources in the
  dataset.  Because the primary data products from these observations
  are the flux density ratios between the sources, the value of the
  flux density assigned to this source is not important.  The only
  rationale for this final step is one of convenience -- to put the
  source visibilities on a scale close to the correct one.
\end{enumerate}

We made no effort to remove any temporal changes in the amplitude
gains, except in the final observing session, which included an
additional 30 point sources.  These were added for the purpose of
investigating the VLA's gain stability over wide ambient temperature
changes.  A very small effect was observed at some frequencies, and
removed with a 6-hour smoothing window.

\subsection{Determination of Flux Density Ratios and Errors}

For each source at each session and frequency, the flux density -- on
the arbitrary scale as noted above -- and a likely error were
determined with the following procedure -- utilized for both
unresolved and resolved objects.
\begin{itemize}
\item The source flux density was determined from an image utilizing
  all the data for one session.  The image provides two estimates, one
  from the sum of the clean components, the other from integration
  over the reconstructed brightness of the source.  For sources which
  were unresolved or marginally resolved, a third estimate from
  a fit to a 2-D Gaussian profile to the image was made.  In all
  cases, these estimates agreed to within the image noise errors.
\item For the four planets, an additional estimation was made from
  fitting the complex visibility to an appropriate limb-darkened disk
  model (\citet{But99}, \citet{But01}).  A comparison of this
  estimate to that from the image total flux showed excellent
  agreement to within the parameter errors generated by the fit.
\item An estimate of the error in the derived mean flux density was
  made from the dispersion of the $N$ individual `snapshot'
  observations of each source at each band.  For each snapshot
  observation, an estimate of the flux density was made by summing the
  clean components generated by the deconvolution.  The dispersion,
  $\sigma$, was determined from these $N$ estimates, and the error
  estimate of the mean, $\sigma_{u}$, was then determined from
  $\sigma_\mu = \sigma/\sqrt{N-1}$.
\end{itemize}

This method for determining the measurement error is valid provided
the errors in the $N$ observations are statistically independent.  We
believe this is the case, since the primary source of error in the
flux determination at the higher frequencies is from residual pointing
errors, which change over a timescale of an hour or less.
Furthermore, any residual errors in the elevation gain dependency
correction are likely due to variable atmospheric opacities, which we
expect are also variable on time-scales of an hour or less.  Since the
observations of any individual source are typically separated by an
hour or more, we can consider each observation independent of the
others.

At low frequencies (90, 20, and 10 cm bands), the errors are dominated
by residuals in the deconvolution procedure since the background
sources are never perfectly removed, and by errors in the corrections
for gain variations.  Again, we believe that the characteristic
timescale for changes in these errors is typically one hour or less,
so that the individual observations are expected to have
independent errors.

The primary data products of this program are the flux density ratios
between all of the sources.  These were determined utilizing the flux
densities derived as described above.  The errors of these ratios were
determined using standard propagation of error methodology, assuming
the errors in the flux density of one source are independent of those
for the others.

\section{Identification of non-variable secondary calibrators}

We seek a set of strong, compact, and non-varying objects which can
serve as general purpose absolute flux density calibrators.  This was
done by examining the changes in the ratios of the flux densities of
our ten non-planetary target objects over the duration of this
project. Since it is extremely unlikely that variability in any one
object is correlated over both time and frequency with variability
with any other, a constant ratio between two sources over the duration
of this project is a strong indicator that both are constant in time.
This procedure has the advantage of being independent of the flux
density scale imposed by the calibration procedure described above.
This analysis was done at the frequencies shown in
Table~\ref{tab:FreqCoverage} at which our observations spanned the
longest period of time: 327.5, 1275, 1465, 4535, 4885, 8435, 8735,
14965, 22460, and 43340 MHz.  For some of the sessions, where a
particular frequency was not utilized, such as in 1995 and 1998, when
1475 MHz was tuned instead of 1465 MHz, the two frequencies actually
tuned within that same band were used, and the desired value derived
by a linear interpolation.

\begin{figure}[ht]
\centerline{\hbox{
\includegraphics[width=6.5in]{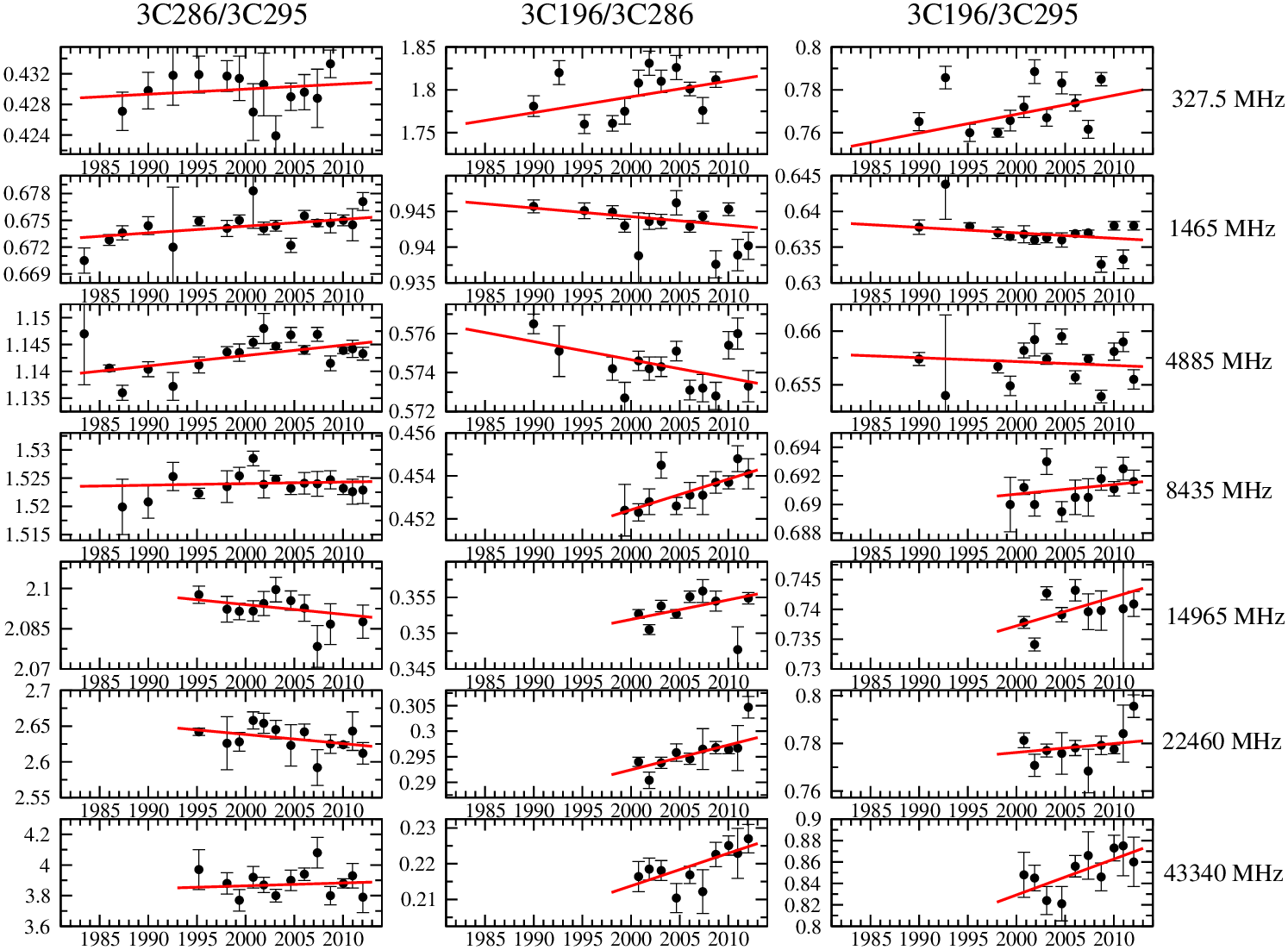}}}
\centerline{\parbox{6in}{
\caption{\small Showing the ratios of the flux densities between
  3C286, 3C196, and 3C295, as a function of time and for selected
  frequencies.  Secular changes are very small, typically a few
  percent/century, except at the highest and lowest frequencies, where
  the errors are highest. The red lines show a weighted linear
  regression fit.}
\label{fig:Steady286}}}
\end{figure}

This examination showed that the four sources 3C123, 3C196, 3C286,
and 3C295, showed negligible changes in their flux density ratios over
the duration of this project, and hence should serve as good absolute
flux density sources. Plots of the ratios of three of these sources
over time are shown in Fig~\ref{fig:Steady286}.  The error bars are
1-$\sigma$ probable errors derived from the estimated errors in the
individual flux densities.  The red lines are from a weighted linear
regression fit.
The slopes of the fits, in percent/century, their errors, and the
reduced $\chi^2$ of the weighted fits are shown in
Table~\ref{tab:Change}.  The fitted values of the flux density ratios
for the year 2000, and the estimated errors of this ratio are shown in
Table~\ref{tab:Ratio}.

\begin{deluxetable}{rrrrrrrrrrrrr}
\tabletypesize{\scriptsize}
\tablecolumns{13}
\tablecaption{Secular Changes in Flux Density Ratios amongst Three Sources
\label{tab:Change}}
\tablewidth{0pt}
\tablehead{
\colhead{Freq.}&
\colhead{286/295}&
\colhead{$\chi^2$}&
\colhead{196/286}&
\colhead{$\chi^2$}&
\colhead{196/295}&
\colhead{$\chi^2$}&
\colhead{123/196}&
\colhead{$\chi^2$}&
\colhead{123/286}&
\colhead{$\chi^2$}&
\colhead{123/295}&
\colhead{$\chi^2$}\\
\colhead{MHz}&
\colhead{\%/century}&\colhead{}&\colhead{\%/century}&\colhead{}&
\colhead{\%/century}&\colhead{}&\colhead{\%/century}&\colhead{}&
\colhead{\%/century}&\colhead{}&\colhead{\%/century}&\colhead{}}
\startdata
327.5&
 1.6$\pm{2.4}$&1.2&10.3$\pm{3.2}$&4.0&11.4$\pm{2.6}$&7.0&-0.1$\pm{2.5}$&1.4&9.8$\pm{2.7}$&5.9&13.3$\pm{2.7}$&9.8\\
1275&
 2.2$\pm{1.7}$&9.8& 1.8$\pm{1.8}$&7.6& 3.2$\pm{1.6}$&0.2&-5.3$\pm{3.2}$&3.3&3.7$\pm{3.5}$&20&-2.4$\pm{3.5}$&3.1\\
1465&
 1.1$\pm{0.3}$&1.6&-1.3$\pm{0.5}$&2.4&-1.2$\pm{0.5}$&3.2&-1.1$\pm{0.9}$&1.3&-0.4$\pm{0.6}$&8.7&0.3$\pm{0.5}$&2.3\\
4535&
-3.2$\pm{1.3}$&1.8&-0.3$\pm{1.3}$&3.2&-3.6$\pm{1.1}$&3.9&-0.8$\pm{1.1}$&6.1&2.2$\pm{1.4}$&4.2&-7.3$\pm{1.0}$&8.7\\
4885&
 1.7$\pm{0.2}$&3.2&-1.6$\pm{0.5}$&3.0&-0.5$\pm{0.7}$&5.5&-0.8$\pm{1.0}$&2.9&0.2$\pm{0.9}$&3.9&-0.3$\pm{0.9}$&7.4\\
8435&
 0.2$\pm{0.4}$&1.5& 3.2$\pm{0.9}$&1.3& 1.0$\pm{0.8}$&1.7&-3.9$\pm{1.1}$&3.5&-1.7$\pm{1.1}$&7.4&-2.8$\pm{0.9}$&7.3\\
8735&
-1.2$\pm{0.8}$&0.5& 3.1$\pm{1.1}$&3.3& 1.2$\pm{0.9}$&1.4&-2.4$\pm{1.1}$&3.7&1.5$\pm{1.0}$&10&-1.5$\pm{1.1}$&6.0\\
14965&
-1.7$\pm{1.4}$&1.1& 7.8$\pm{2.2}$&3.0& 6.6$\pm{2.3}$&4.4&-12 $\pm{3.1}$&2.7&-2.8$\pm{3.3}$&2.2&-6.0$\pm{3.2}$&3.3\\
22460&
-4.8$\pm{1.5}$&1.1&  17$\pm{3.6}$&2.1& 4.8$\pm{3.8}$&2.2&-11 $\pm{5.6}$&4.4&6.5$\pm{5.9}$&3.7&-7.5$\pm{6.5}$&1.5\\
43340&
 4.8$\pm{9.5}$&1.4&  43$\pm{15}$ &1.1&  40$\pm{17}$ &0.9&-61 $\pm{94} $&2.2&-11$\pm{102}$&0.1&-41$\pm{102}$&0.0\\
\enddata

\tablecomments{{Listed are the linear rate of change in flux density
  ratio amongst three of the four apparently constant sources, in
  percent/century, along with the estimated error of this slope
  and reduced $\chi^2$ of a weighted fit.}}
\end{deluxetable}

\begin{deluxetable}{lllllll}
\tabletypesize{\scriptsize} 
\tablecaption{Flux Density Ratios for date 2000.0 for the Four Primary Sources
\label{tab:Ratio}}
\tablewidth{0pt} 
\tablehead{
\colhead{Freq.}&
\colhead{R(286/295)}&
\colhead{R(196/286)}&
\colhead{R(196/295)}&
\colhead{R(123/196)}&
\colhead{R(123/286)}&
\colhead{R(123/295)}}
\startdata
327.5&
0.4299$\pm{0.0007}$&1.7919$\pm{0.0034}$&0.7686$\pm{0.0011}$&
3.1012$\pm{0.0050}$&5.5688$\pm{0.0100}$&2.3990$\pm{0.0035}$\\
1275&
0.6435$\pm{0.0006}$&0.9943$\pm{0.0010}$&0.6399$\pm{0.0006}$&
3.3565$\pm{0.0043}$&3.3215$\pm{0.0050}$&2.1482$\pm{0.0031}$\\
1465&
0.6743$\pm{0.0002}$&0.9443$\pm{0.0003}$&0.6370$\pm{0.0002}$&
3.3872$\pm{0.0014}$&3.1923$\pm{0.0010}$&2.1554$\pm{0.0006}$\\
4535&
1.1041$\pm{0.0008}$&0.5929$\pm{0.0004}$&0.6541$\pm{0.0004}$&
3.7671$\pm{0.0023}$&2.2339$\pm{0.0016}$&2.4688$\pm{0.0014}$\\
4885&
1.1430$\pm{0.0002}$&0.5747$\pm{0.0002}$&0.6572$\pm{0.0003}$&
3.7945$\pm{0.0019}$&2.1772$\pm{0.0010}$&2.4918$\pm{0.0012}$\\
8435&
1.5240$\pm{0.0004}$&0.4524$\pm{0.0003}$&0.6907$\pm{0.0004}$&
4.0188$\pm{0.0026}$&1.8210$\pm{0.0013}$&2.7782$\pm{0.0017}$\\
8735&
1.5551$\pm{0.0009}$&0.4456$\pm{0.0003}$&0.6935$\pm{0.0004}$&
4.0288$\pm{0.0023}$&1.7944$\pm{0.0009}$&2.7944$\pm{0.0017}$\\
14965&
2.0940$\pm{0.0014}$&0.3519$\pm{0.0004}$&0.7372$\pm{0.0008}$&
4.2670$\pm{0.0052}$&1.5022$\pm{0.0024}$&3.1496$\pm{0.0041}$\\
22460&
2.6383$\pm{0.0031}$&0.2924$\pm{0.0007}$&0.7762$\pm{0.0023}$&
4.4526$\pm{0.0139}$&1.3033$\pm{0.0046}$&3.4551$\pm{0.0026}$\\
43340&
3.864 $\pm{0.026} $&0.2138$\pm{0.0022}$&0.829 $\pm{0.010}$&
4.5   $\pm{0.5}$   &0.93$\pm{0.10}$&3.7$\pm{0.4}$\\
\enddata

\tablecomments{The ratio values at 43.3 GHz involving 3C123 are of
much poorer quality due to the large anglar size of that source.}

\end{deluxetable}

The results of this analysis show that secular changes in the flux
densities of these four objects are less than 1\% over the 30-year
duration of this project, with the exception of the highest
frequencies, where the errors limit the accuracy to a few times higher
than this.  

The unchanging ratio between 3C286 and 3C295 is physically reasonable.
VLBI observations of 3C286 show that there is no identifiable inverted
spectrum core, and VLBA and EVN polarimetric images (\citet{Cot97b},
\citet{Jia96}) show a uniform polarization throughout the source's
structure.  With no visible compact core, and no detected source
expansion, time variability must be of order the light-travel time
across the source -- hundreds of years for 3C286.  As discussed in
\citet{Ott94}, no variation in 3C295 is expected due to its large
physical size ($\sim$ 5 kpc) and absence of a significant nuclear core
(\citet{Tay92}).  

\section{Converting Ratios to Absolute Flux Densities}

The procedures described above provided ratios between the observed
sources accurate to better than 1\% at all but the highest frequency
band.  Conversion of these ratios to absolute flux densities requires
at least one source whose absolute flux density is known.  We utilize
emission models of the planet Mars, combined with absolute
measurements of its emission from WMAP, for this purpose, following
the procedures described below.

\subsection{The Rudy Model Fit to WMAP observations}

The flux density of Mars is variable as a function of time, due to
both geometrical factors (distance, and subearth latitude and
longitude) and to seasonal factors (because of the waxing and waning
of the polar caps).  It is thus necessary to use a thermophysical
model to calculate the expected emission from the planet.  Such
thermophysical models exist, for example those of \citet{Wri76} and
\citet{Rud87}. We choose to use the latter model (\citet{Rud87},
\citet{Muh91}, \citet{Sid00}), because it is based on observations at
centimeter wavelengths which overlap those of the WMAP measurements,
which are absolute (\citet{Wei11}), as they are calibrated
against the CMB dipole.

WMAP measured the brightness temperature of Mars at five frequencies
(22.85 GHz (K-band); 33.11 GHz (Ka-band); 40.82 GHz (Q-band); 60.85
GHz (V-band); and at 93.32 GHz (W-band)) during seven `seasons' of
observing (\citet{Wei11}).  Three of these frequency bands overlap
those available at the VLA, thus providing a good way to transfer the
WMAP flux density scale to the higher frequencies of the VLA.  By
utilizing the thermophysical emission model described above, and
incorporating lower-frequency observations described in \citet{Baa77},
the entire radio frequency spectrum from $\sim$1 GHz to 50 GHz can be
placed on an absolute calibration scale.

The spectral flux density measured by a total power telescope such as
WMAP for any particular pointing is the antenna beam-weighted integral
of the sky brightness over the full sky:
\begin{equation}
S = \int BPd\Omega
\end{equation}
where $B$ is the sky brightness, $P$ is the normalized antenna power
pattern, and the integration is over $4\pi$ steradians.  The {\sl
increment} in spectral flux density, $S_p$, provided by a uniform, discrete
source of solid angle $\Omega_p$ (assumed much smaller than the
antenna main beam), brightness $B_p$ and opacity $\tau$, -- hence the
value actually measured by the telescope -- is
\begin{equation}
S_p = [B_p-B_{bg}(1-e^{-\tau})]\Omega_p.
\end{equation}
where $B_{bg}$ is the sky brightness behind the foreground object.  The
second term on the right-hand side accounts for the attenuation of the
background sky by the source.  For an optically thick source, such as a
planet, this becomes
\begin{equation}
S_p = (B_p - B_{bg})\Omega_p.
\end{equation}
Thus, flux density measured by the telescope is proportional to the
brightness difference between the object and the background sky. To
this value must be added that amount occulted by the object in order
to determine the actual flux density.  These considerations apply
equally to interferometers, as the uniform background sky is resolved
out by the interferometric coherence pattern.

For our application, both the planet brightness $B_p$ and the
background sky brightness $B_{bg}$ are described by the Planck
function
\begin{equation}
B(\nu,T) = \frac{2h\nu^3}{c^2}\frac{1}{e^{h\nu/kT}-1}
\end{equation}
where $T$ is the brightness temperature of the source.  The WMAP
brightness temperatures of Mars published in \citet{Wei11} are given in terms
of the `Rayleigh-Jeans' brightness temperature, defined as
\begin{equation}
T_{RJ} = \frac{\lambda^2 S_\nu}{2k\Omega},
\end{equation}
and have not been adjusted by the blockage of the sky background.
Since the Rudy model provides an estimate of the total brightness of
the source in true `Planck' units, we must adjust the published WMAP
values for both the blockage and the difference in units.

>From equations 5, 6, and 7, we can relate the true brightness
temperature of an optically thick object, $T_P$, to the value assigned
using the Rayleigh-Jeans approximation, $T_{RJ}$, accounting for the
flux blocked by the planet's disk:
\begin{equation}
\frac{h\nu}{kT_P}=\ln\left[1+\frac{1}{(e^{h\nu/kT_{bg}}-1)^{-1}
+\frac{kT_{RJ}}{h\nu}}\right]
\end{equation}
With $T_{bg}=2.75K$, we find that the published WMAP values for Mars
must be increased by 2.78, 2.82, 2.86, 3.00, and 3.32 K for the five
frequency bands used by WMAP.  The WMAP observations of the brightness
temperature of Mars for all frequencies and sessions are shown in
Table~\ref{tab:MarsTemp}, along with the values adjusted as described
above.

\begin{deluxetable}{cccccccccccccccc}
\tabletypesize{\scriptsize} 
\tablecaption{WMAP Observations and Rudy
Model Data\label{tab:MarsTemp}} 
\tablewidth{0pt} 
\tablehead{
\colhead{JD}&
\multicolumn{3}{c}{22.85 GHz}&
\multicolumn{3}{c}{33.11 GHz}&
\multicolumn{3}{c}{40.82 GHz}&
\multicolumn{3}{c}{60.85 GHz}&
\multicolumn{3}{c}{93.32 GHz}\\
\colhead{}&
\colhead{W}&\colhead{C}&\colhead{M}&
\colhead{W}&\colhead{C}&\colhead{M}&
\colhead{W}&\colhead{C}&\colhead{M}&
\colhead{W}&\colhead{C}&\colhead{M}&
\colhead{W}&\colhead{C}&\colhead{M}}

\startdata 
2182.62&178$\pm$4&181&195&182$\pm$3&185&195&186$\pm$4&189&195&191$\pm$3&194&196&189$\pm$2&192&196\\ 
2776.39&183$\pm$4&186&194&187$\pm$3&190&196&191$\pm$4&194&198&197$\pm$3&200&202&204$\pm$2&207&206\\ 
2983.75&191$\pm$4&194&200&195$\pm$3&198&200&185$\pm$4&188&201&193$\pm$3&196&201&195$\pm$2&198&202\\ 
3586.17&200$\pm$3&203&207&199$\pm$2&202&210&203$\pm$3&205&212&209$\pm$2&212&215&213$\pm$1&216&220\\ 
3758.26&191$\pm$5&194&190&184$\pm$3&187&190&189$\pm$4&192&190&186$\pm$3&189&190&185$\pm$2&188&192\\ 
4389.29&186$\pm$4&189&196&187$\pm$3&190&198&196$\pm$4&199&199&197$\pm$3&200&203&198$\pm$2&201&207\\ 
4530.49&174$\pm$6&177&184&177$\pm$4&180&184&176$\pm$5&179&184&181$\pm$4&184&185&182$\pm$2&185&186\\ 
\enddata 

\tablecomments{For each WMAP observing band is listed the published WMAP
brightness temperature (`W'), the temperature after correction for the sky
background and conversion to Planck units (`C'), and the Rudy model
prediction (`M').  The listed JD date has had 2450000 subtracted.}
\end{deluxetable}
   
\begin{figure}[ht!]
\centerline{\hbox{
\includegraphics[height=6in,angle=-90,origin=c]{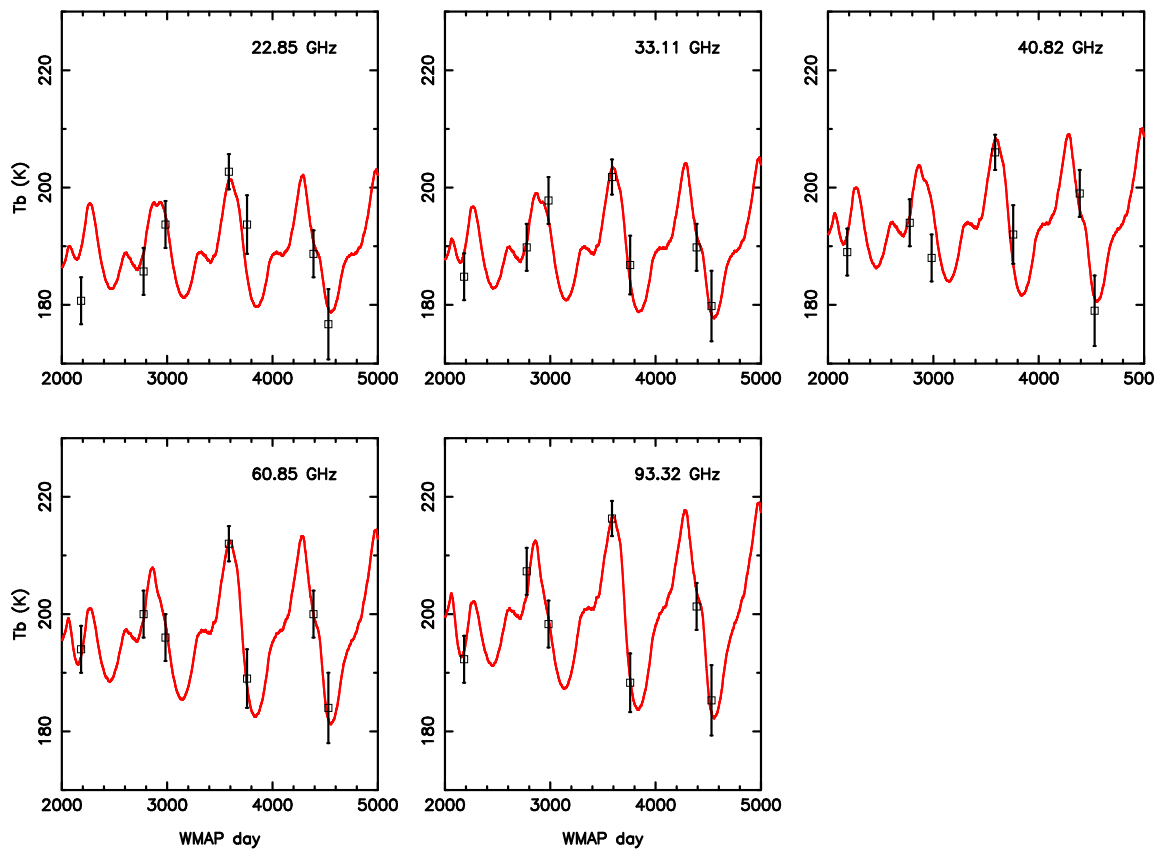}}}
\centerline{\parbox{6in}{
\caption{\small The red line shows the predictions of the adjusted
  Rudy model to the WMAP observations, shown as the black data points.
  The regular oscillations are due to a combination of orbital motions
  of the Earth and Mars, the inclination of Mars, and Martian seasonal
  variations.}
\label{fig:RudyWMAPFits}}}
\end{figure}

Given the WMAP observation dates, we then calculate the Rudy Model
brightness temperatures for each observation, averaged over a full
rotation of Mars, since the Rudy Model has variations as a function of
planetary longitude, based on surface thermophysical properties, and
the \citet{Wei11} results do not include a time of day of the
observations.  The Rudy Model values calculated in this way are also
shown in Table~\ref{tab:MarsTemp}.  Figure~\ref{fig:RudyWMAPFits}
shows the fits along with the model calculations for all five bands.

With the WMAP observations and the Rudy Model predictions for a given
frequency, we then find a correction factor to apply to the Rudy Model
to make the best agreement with the WMAP observations, given by:
\begin{equation}
f = \frac{\sum w_iM_iD_i}{\sum w_iM_i^2}
\end{equation}
where the $M_i$ are the Rudy Model values, $D_i$ the WMAP observed
values, and $w_i$ the WMAP weights ($w_i = 1/\sigma^2$).  

Table~\ref{tab:CorrFact} shows the derived corrections at each band,
and an error estimated from the dispersion in the individual ratios --
for these estimates we did not use the `season 1' WMAP observations in
this fit, since they were taken during a global dust storm
(\citet{Wei11}).  These storms are known to change the planet's
emission, and are not included in the model.
\begin{deluxetable}{cccccc}
\tabletypesize{\scriptsize}
\tablecaption{Rudy Model Correction Factors for Agreement with WMAP
  Data
\label{tab:CorrFact}}
\tablewidth{0pt}
\tablehead{
\colhead{} & \colhead{K}&\colhead{Ka}&\colhead{Q}&\colhead{V}&\colhead{W}}
\startdata
 Ratio&  0.974&   0.970&   0.985&   0.987&   0.987\\
 Error&  0.008&   0.005&   0.011&   0.003&   0.005\\
\enddata
\end{deluxetable}  

We find little evidence of the unmodeled frequency dependent
emissivity reported by \citet{Wei11}; what they reported may have been
a byproduct of their extrapolation of the Wright Model to
inappropriate frequencies.  Because we find no strong evidence for
variation of the correction factor across these five bands, we have
applied the average correction factor of 0.975 from the lowest three
WMAP frequencies to the spectral flux densities derived from the Rudy
model.

\subsection{Generating the Apparent Flux Density of Mars}

The Rudy model calculates the {\sl total} spectral flux density from
the planet Mars.  To determine that actually measured by the VLA
requires subtracting that portion blocked by the planet's disk, as
described in Section 7.1.  The background brightness temperature $T_{bg}$ is
\begin{equation}
      T_{bg} = T_{gal} + T_{cmb}.
\end{equation}
where $T_{cmb} = 2.725$ K and $T_{gal}$ is the brightness temperature
of the galactic background.  This is widely variable, depending on
galactic longitude and latitude.  For our observations, Mars was
always sufficiently far from the galactic plane that we can utilize
the expression $T_{gal}\sim 2.5\nu_G^{-2.7 }$ K, where $\nu_G$ is the
frequency in GHz, to estimate the galactic contribution.  In general,
the galactic synchrotron brightness is negligible compared to the CMB
at frequencies above ~2 GHz.  For observations below 2 GHz, the
galactic background can contribute a small offset, but this is
completely negligible compared to the errors in the observations due
to the low flux density of Mars below 2 GHz.  

The results of this procedure are shown in Table~\ref{tab:MarsFlux},
giving the apparent flux density of Mars for the dates, times, and
frequencies shown.  Also shown are the approximate coordinates, in
equatorial and galactic coordinates, of Mars during our observations.
\begin{deluxetable}{lcrrrcrcllllll}
\tabletypesize{\scriptsize} 
\tablecaption{Calculated Apparent Mars
Spectral Flux Densities at Selected Frequencies
\label{tab:MarsFlux}}
\tablewidth{0pt} 
\tablehead{
\colhead{Date}&
\colhead{$\alpha$}&\colhead{$\delta$}&\colhead{\sl{l}}&\colhead{\sl{m}}&
\colhead{Dist.}&\colhead{Diam.}&\colhead{1465}&
\colhead{4885}&\colhead{8435}&\colhead{14965}&\colhead{22460}&
\colhead{43340}\\
\colhead{}&\colhead{hh mm}&\colhead{sdd mm}&\colhead{ddd}&\colhead{sdd}&
\colhead{AU}&\colhead{arcsec.}&\colhead{MHz}&\colhead{MHz}&
\colhead{MHz}&\colhead{MHz}&\colhead{MHz}&\colhead{MHz}
} 
\startdata 
1995 Mar 14&09 10& 20 18&208& 39&0.77&12.1&     &      &      &      & 7.47&27.7\\
1998 Feb 03&22 37&-09 44& 55&-54&2.24& 4.2&     &      &      &0.428 &0.973&3.68\\ 
1999 Apr 16&14 21&-12 19&335& 45&0.60&15.6&.0531&0.599 & 1.79 &5.67  & 12.8&47.9\\ 
2000 Oct 03&10 49& 08 52&239& 56&2.45& 3.8&     &0.0356&0.107 &0.335 &0.765&2.89\\ 
2001 Nov 10&20 50&-19 39& 27&-35&1.13& 8.3&     &0.173 &0.521 &1.65  &3.72 &13.8\\ 
2003 Feb 07&16 49&-22 08&358& 14&1.72& 5.4&     &0.0723&0.216 &0.686 &1.56 &5.92\\ 
2004 Aug 27&10 50& 08 33&240& 56&2.66& 3.5&     &0.0301&0.0901&0.285 &0.646&2.43\\ 
2006 Jan 17&00 52& 18 16&123&-45&0.92&10.2&.0234&0.263 &0.771 &2.414 &5.40 &20.0\\ 
2007 May 14&23 59&-01 43& 94&-62&1.71& 5.5&     &0.0747&0.227 & 0.726&1.66 &6.32\\ 
2008 Sep 14&13 02&-06 15&308& 56&2.45& 3.8&     &0.0355&0.106 & 0.336&0.773&2.85\\ 
2010 Jan 11&09 20& 19 55&210& 41&0.69&13.5&.0395&0.442 & 1.32 &      &9.43 &35.3\\ 
2010 Dec 26&19 01&-23 39& 13&-13&2.38& 3.9&     &0.0384&0.116 &0.368 &0.839&3.17\\ 
2012 Jan 19&11 39& 06 03&260& 63&0.88&10.6&.0243&0.272 &0.812 &2.57  &5.82 &21.9\\ 
\enddata

\tablecomments{The spectral flux densities in Jy of Mars, adjusted for
background sky brightness, visible to the VLA.  Values listed are
averages over each session.  Missing entries correspond to dates when
Mars was not observed, or not detected.  Values listed have been
adjusted for the attenuation due to the VLA's primary beam.}

\end{deluxetable}
For each observing session and frequency, the ratio between our
observed spectral flux density of Mars, based on the arbitrary
calibration scale described earlier, and the calculated apparent
spectral flux density of Mars shown in Table~\ref{tab:MarsFlux} was
used to adjust the derived flux densities of the four non-varying
sources 3C123, 3C196, 3C286, and 3C295 to the Mars-based scale.  For
the 90cm band ratios, where the VLA cannot detect Mars, we have used
the \citet{Sca12} value for 3C196 as the reference source.

\section{Polynomial Expressions for the Flux Densities of 3C123, 3C196, 3C286, and 3C295}

Our derived flux densities for each of the four primary flux density
calibrators, based on their ratios to the planet Mars and utilizing
the Rudy Model adjusted to the WMAP scale are given in
Table~\ref{tab:StdFlux}.  The values listed at 327.5 MHz -- where Mars
is too weak to be detected -- are based on our measured ratios to the
source 3C196, whose spectral flux density is set to the
\citet{Sca12}value of 46.75 Jy.  We emphasize that the \citet{Sca12}
scale is not independently based on an absolute standard, but utilizes
various published observations from the literature which are then tied
to a proposed common flux density scale.  We include these low
frequency values to our fits in order to provide a plausible
low-frequency extension of our proposed scale.

The listed errors shown in Table~\ref{tab:StdFlux} are estimated from the
dispersion in the measurements of the individual ratios to Mars taken
since 1995.  The number of separate observations for each band is
given in the right-hand most column.
\begin{deluxetable}{cccccccccr}
\tabletypesize{\scriptsize}
\tablecaption{Adopted Spectral Flux Densities of Steady Sources 
\label{tab:StdFlux}}
\tablewidth{0pt}
\tablehead{
\colhead{Freq.}&\multicolumn{2}{c}{3C123}&\multicolumn{2}{c}{3C196}&
\multicolumn{2}{c}{3C286}&\multicolumn{2}{c}{3C295}&$N_{obs}$\\
\colhead{(MHz)}&\colhead{S(Jy)}&\colhead{$\sigma$(Jy)}&\colhead{S(Jy)}&
\colhead{$\sigma$(Jy)}&\colhead{S(Jy)}&\colhead{$\sigma$(Jy)}&
\colhead{S(Jy)}&\colhead{$\sigma$(Jy)}&}
\startdata
0.3275&145.0&4.3 &46.8 &1.4 &26.1 &0.8 &60.8 &1.8&14\\
1.015 &66.2 &4.3 &20.1 &4.8 &18.4 &4.3 &30.8 &7.3&1\\
1.275 &46.6 &3.2 &13.3 &2.0 &13.8 &2.0 &21.5 &3.0&1\\
1.465 &47.8 &0.5 &14.1 &0.2 &15.0 &0.2 &22.2 &0.5&4\\
1.865 &38.7 &0.6 &11.3 &0.2 &13.2 &0.2 &17.9 &0.3&2\\
2.565 &28.9 &0.3 &8.16 &0.1 &10.9 &0.2 &12.8 &0.2&2\\
3.565 &21.4 &0.8 &6.22 &0.2 & 9.5 &0.1 &9.62 &0.2&3\\
4.535 &16.9 &0.2 &4.55 &.06 &7.68 &0.1 &6.96 &.09&7\\
4.835 &16.0 &0.2 &4.22 &.1  &7.33 &0.2 &6.45 &.15&1\\
4.885 &15.88&0.1 &4.189&.025&7.297&.046&6.37 &.04&11\\
6.135 &12.81&.15 &3.318&.05 &6.49 &.15 &4.99 &.05&2\\
6.885 &11.20&.14 &2.85 &.05 &5.75 &.05 &4.21 &.05&3\\
7.465 &11.01&.2  &2.79 &.05 &5.70 &.10 &4.13 &.07&1\\
8.435 &9.20 &.04 &2.294&.010&5.059&.021&3.319&.014&11\\
8.485 &9.10 &.15 &2.275&.03 &5.045&.07 &3.295&.05&1\\
8.735 &8.86 &.05 &2.202&.011&4.930&.024&3.173&.016&10\\
11.06 &6.73 &.15 &1.64 &.03 &4.053&0.08&2.204&.05&1\\
12.890&     &    &1.388&.025&3.662&.070&1.904&.04&1\\
14.635&5.34 &.05 &1.255&.020&3.509&.040&1.694&.04&1\\
14.715&5.02 &.05 &1.206&.020&3.375&.040&1.630&.03&1\\
14.915&5.132&.025&1.207&.004&3.399&.016&1.626&.008&7\\
14.965&5.092&.028&1.198&.007&3.387&.015&1.617&.007&11\\
17.422&4.272&.07 &0.988&.02 &2.980&.04 &1.311&.025&1\\
18.230&     &    &0.932&.020&2.860&.045&1.222&.05&1\\
18.485&4.090&.055&0.947&.015&2.925&.045&1.256&.020&1\\
18.585&3.934&.055&0.926&.015&2.880&.04 &1.221&.015&1\\
20.485&3.586&.055&0.820&.010&2.731&.05 &1.089&.015&1\\
22.460&3.297&.022&0.745&.003&2.505&.016&0.952&.005&13\\
22.835&3.334&.06 &0.760&.010&2.562&.05 &0.967&.015&1\\
24.450&2.867&.03 &0.657&.017&2.387&.03 &0.861&.020&2\\
25.836&2.697&.06 &0.620&.017&2.181&.06 &0.770&.02 &1\\
26.485&2.716&.05 &0.607&.017&2.247&.05 &0.779&.020&1\\
28.450&2.436&.06 &0.568&.015&2.079&.05 &0.689&.020&2\\
29.735&2.453&.05 &0.529&.015&2.011&.05 &0.653&.020&1\\
36.435&1.841&.17 &0.408&.005&1.684&.02 &0.484&.015&3\\
43.065&     &    &0.367&.015&1.658&.08 &0.442&.020&1\\
43.340&1.421&.055&0.342&.005&1.543&.024&0.398&.006&13\\
48.350&1.269&.12 &0.289&.005&1.449&.04 &0.359&.013&4\\
48.565&     &    &0.272&.015&1.465&.1  &0.325&.025&1\\
\enddata

\tablecomments{The derived flux density values, based on the Mars
  emission model for frequencies between 1 and 50 GHz.  The quoted
  errors are derived from the dispersion of the values over the
  various sessions.  The values for 3C123, 3C286, and 3C295 at 327.5
  MHz are derived from their ratios to 3C196, whose flux density is
  taken to be 46.8 Jy (\citet{Sca12}).}

\end{deluxetable}

\begin{figure}[ht]
\centerline{\hbox{
\includegraphics[width=6.25in]{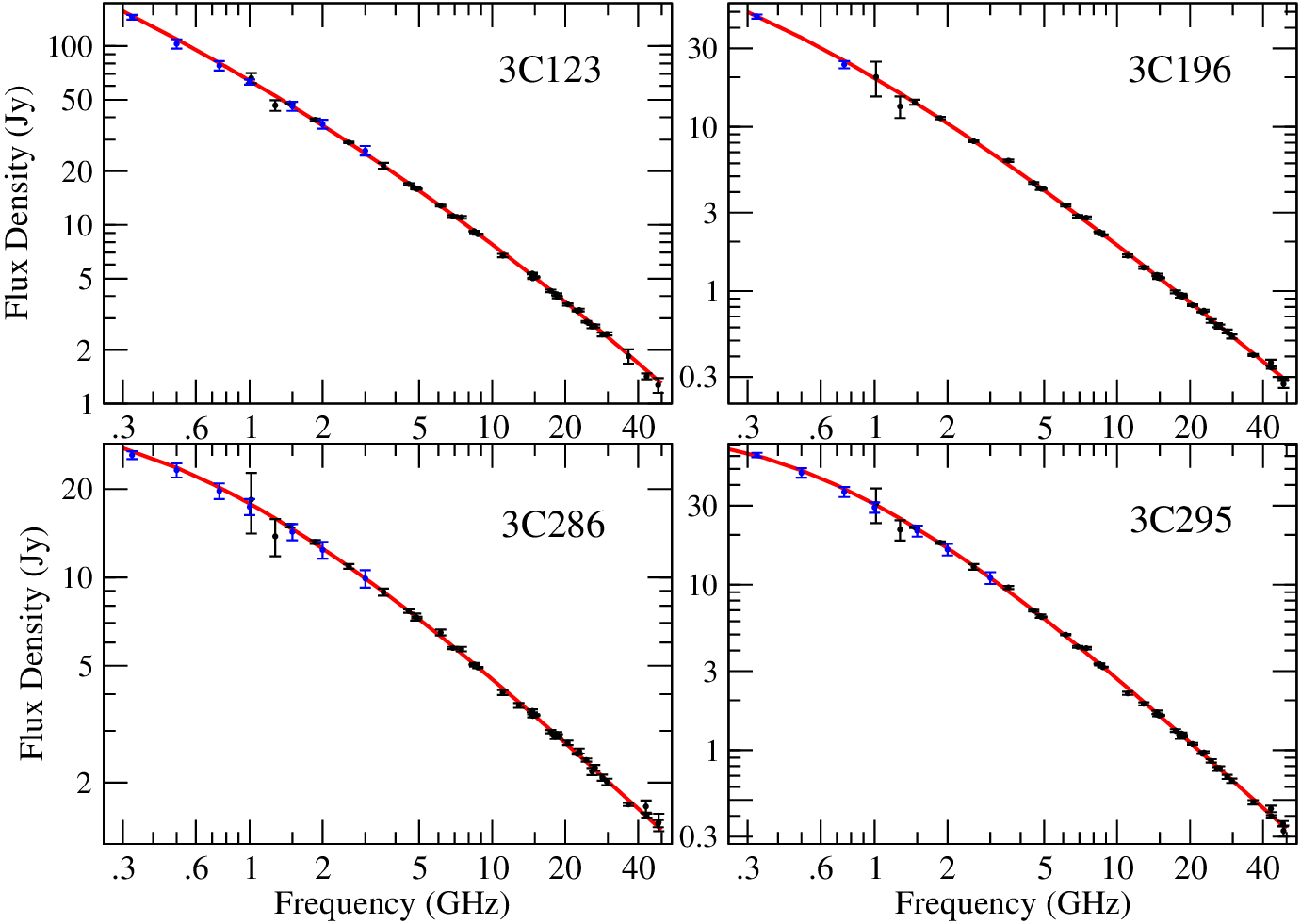}}}
\centerline{\parbox{6.25in}{
\caption{\small Plotted are the measurements of the flux densities,
and their 1-$\sigma$ errors for the four non-variable sources, and the
best-fit models for each.  Data based on the Mars model are plotted in
black.  Data based on the Baars scale, or the \citet{Sca12} scale are
plotted in blue.}
\label{fig:FourPlot}}}
\end{figure}

The frequency dependencies of the spectral flux densities for the four
stable sources were modelled with a cubic polynomial function of the
form
\begin{equation}
\log(S)=a_0 + a_1\log(\nu_G) + a_2[\log(\nu_G)]^2 + a_3[\log(\nu_G)]^3
\end{equation}
where $S$ is the spectral flux density in Jy, and $\nu_G$ is the
frequency in GHz.  Because our values for frequencies below 4 GHz have
much higher errors, and to enable a plausible low-frequency extension
of our scale, we have added the \citet{Baa77} values for 3C123, 3C286,
and 3C295 at 0.5, 0.75, 1.0, 2.0, and 3.0 GHz, with the errors given
in their paper.  The resulting coefficients are given in
Table~\ref{tab:FitFlux}, along with the estimated errors of these
coefficients.  The adopted spectral flux densities, including the
Baars values, and the derived fits, for the four steady objects are
shown in Fig.~\ref{fig:FourPlot}.
\begin{deluxetable}{llllll}
\tabletypesize{\scriptsize}
\tablecaption{Fitted Coefficients for the Four Steady Sources
\label{tab:FitFlux}}
\tablewidth{0pt}
\tablehead{
\colhead{Source}&\colhead{$a_0$}&\colhead{$a_1$}&\colhead{$a_2$}&
\colhead{$a_3$}&\colhead{$\chi^2$}}
\startdata
3C123&1.8077$\pm$0.0036&-0.8018$\pm$0.0081&-0.1157$\pm$0.0047&      0          &2.01\\
3C196&1.2969$\pm$0.0040&-0.8690$\pm$0.0114&-0.1788$\pm$0.0150&0.0305$\pm$0.0063&2.09\\
3C286&1.2515$\pm$0.0048&-0.4605$\pm$0.0163&-0.1715$\pm$0.0208&0.0336$\pm$0.0082&1.22\\
3C295&1.4866$\pm$0.0036&-0.7871$\pm$0.0110&-0.3440$\pm$0.0160&0.0749$\pm$0.0070&1.65\\
\enddata
\end{deluxetable}

Amongst the four sources, 3C286 is the most compact and has the
flattest spectral index, making it the preferred primary flux density
reference source.  We thus adopt 3C286 as our reference source, and
determine the flux densities of all other sources in our program from
their ratios to it.

\section{Estimated Error in the Scale}

There are several sources of error in the measurement and transfer
process.  Here we quantify these. 

\begin{enumerate}
\item The intrinsic error in the WMAP brightness scale.  This is based
  on the knowledge of the CMB monopole, given by \citet{Jar11} as
  0.2\%. 
\item The error in the correction factor which adjusts the Rudy model
  to the WMAP scale.  We estimate this from the
  dispersion in the ratios between the Rudy model predictions and
  the actual WMAP observations.  The dispersion in this ratio for the
  18 separate observations at 22.85, 33.11, and 40.82 GHz (excluding
  the first epoch) is 0.019.  Presuming each is independent, the
  error in the mean is 0.5\%.    
\item An additional error at low frequencies, where we cannot use WMAP
  observations to directly correct the Rudy model, but rather rely on
  the physics of the model properly accounting for the frequency
  dependence of the emission.  Fortunately, the  Rudy model is best
  constrained at those frequencies, since it is based on VLA
  observations at 5 and 15 GHz.  We estimate this error as 1\%, based
  on runs of the model with varying physical parameters, and
  consideration of unmodeled effects (see also the discussion in
  \citet{Muh91}.
\item The error in the transfer from the Mars flux density to that of
  our chosen standard source, 3C286. This is given by the error in our
  ratio measurements, as shown in Table~\ref{tab:StdFlux}.  We
  estimate this error as 2\% at 1465 MHz, 0.7\% at 4885 MHz, 0.5\% at
  8435 and 14965 MHz, rising to 1\% at 22460 MHz, and 2\% at 43340
  MHz.     
\end{enumerate}

Presuming that these four error sources are independent, the total
error is given by adding all four in quadrature -- roughly 2.5\% at
1500 MHz, 1\% from 4000 through 15000 MHz, 2\% at 22000 MHz, and 3\%
at 43000 GHz.

\section{Spectral Flux Densities and Structures of the Remaining Target Sources}

The adoption of 3C286 as the primary flux density standard permits
derivation of the spectral flux densities, as a function of frequency
and time, for the remaining sources in our program.  Here we show the
results for the non-planetary target sources.

\subsection{3C48}

\begin{figure}[ht]
\centerline{{\includegraphics[width=2.4in]{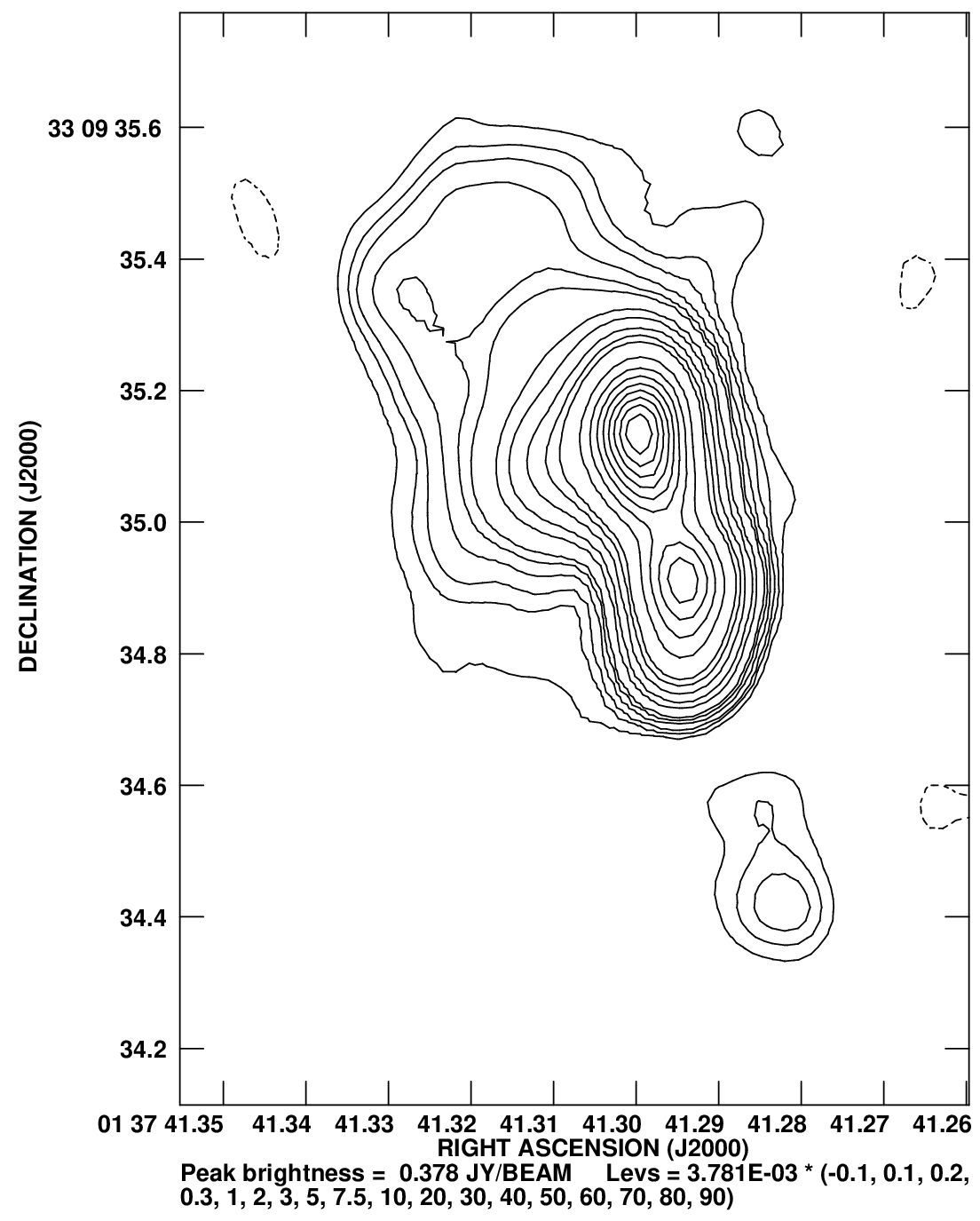}}
\hspace{3mm}{\includegraphics[width=3.7in]{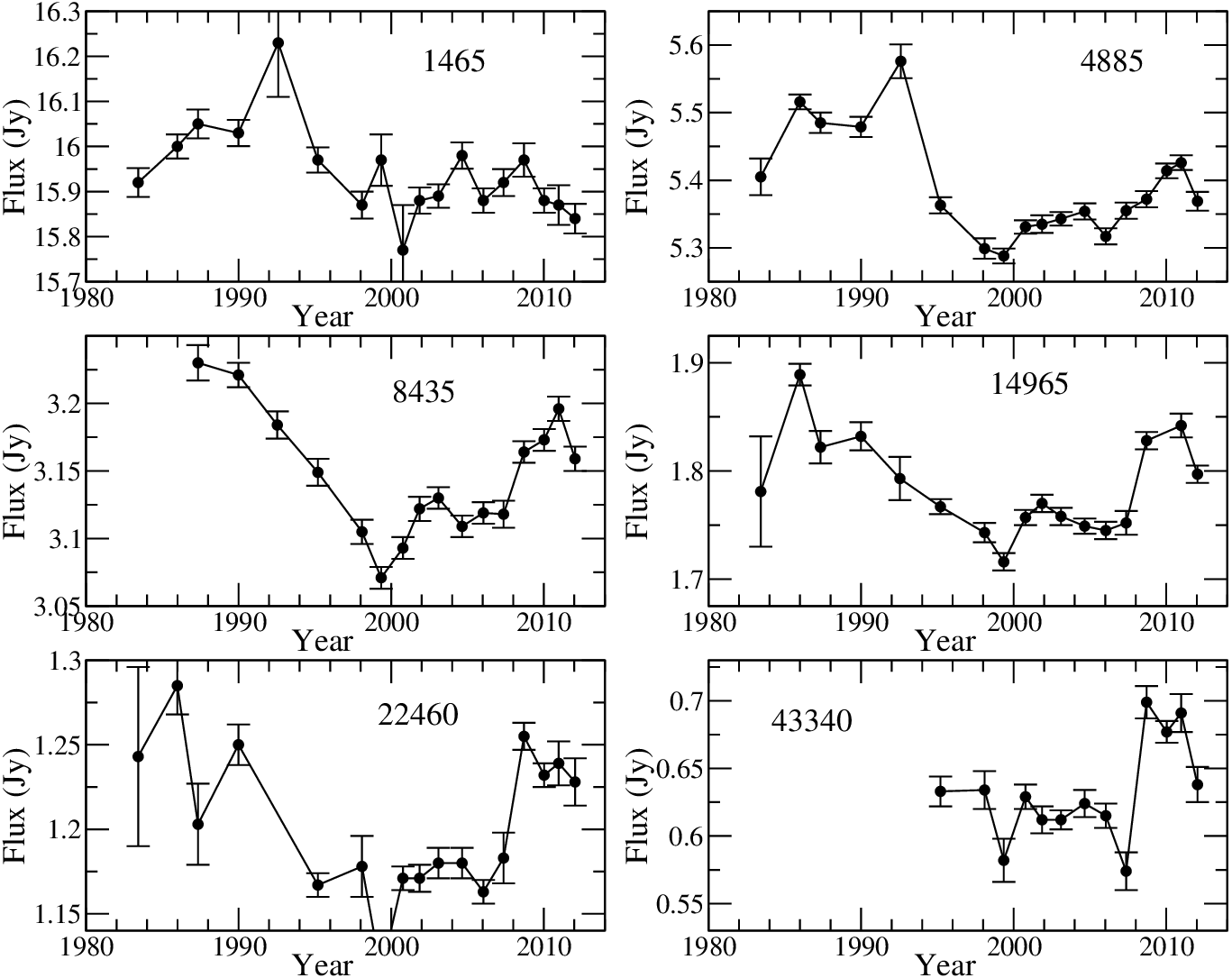}}}
\centerline{\parbox{6in}{
\caption{\small The left panel shows the structure of 3C48 at 22.5 GHz,
  with 100 milliarcsecond resolution.  The active nucleus of the
  quasar is identified with the lower of the two bright peaks.  The
  right panel shows the temporal evolution of the source flux density.}
\label{fig:3C48Maps}}}
\end{figure}

3C48 is a compact steep-spectrum quasar at redshift z=0.367.  Its
radio structure is quite compact with a maximum extent of about 1.2
arcseconds, as shown by the 0.1 arcsecond resolution VLA image shown
in Fig.~\ref{fig:3C48Maps}.  The brightest regions of this source have
been imaged by VLBA, MERLIN, and EVN observations (\citet{An09}).  
The source is a commonly-used flux density calibration source for the
VLA.  The plots in the right panel of Fig.~\ref{fig:3C48Maps} show
that the source has undergone significant changes in flux density over
the past 30 years.  A period of slow decline or quiesence was followed
by a small flare, more prominent at the higher frequencies, beginning
in 2009, which is now declining.  As this source is a commonly utilized
flux density calibration source, we have fitted its changing spectrum
with a cubic polynomial fit for each of the 18 sessions.  The results
of this are given in Table~\ref{tab:CalTimeSpec}.
\begin{deluxetable}{lllllllllllll}
\tabletypesize{\scriptsize}
\tablecaption{Fitted Coefficients for Temporal Change for 3C48, 3C138,
  and 3C147 
\label{tab:CalTimeSpec}}
\tablewidth{0pt}
\tablehead{
\colhead{Session}&\multicolumn{4}{c}{3C48}&\multicolumn{4}{c}{3C138}&
\multicolumn{4}{c}{3C147}\\
&\colhead{A0}&\colhead{A1}&\colhead{A2}&\colhead{A3}&
\colhead{A0}&\colhead{A1}&\colhead{A2}&\colhead{A3}&
\colhead{A0}&\colhead{A1}&\colhead{A2}&\colhead{A3}}
\startdata
1983.4&1.3339&-.7643&-.1946&.055&1.0328&-.5523&-.1161& .008&1.4620&-.7085&-.2347&.051\\
1985.9&1.3350&-.7598&-.1869&.057&1.0337&-.5591&-.1605& .032&1.4648&-.7177&-.2501&.089\\
1987.3&1.3361&-.7577&-.1905&.048&1.0354&-.5914&-.1032&-.005&1.4624&-.7115&-.2336&.071\\
1989.9&1.3363&-.7605&-.1965&.057&1.0292&-.5636&-.1857& .052&1.4646&-.7194&-.2532&.092\\
1995.2&1.3359&-.7673&-.2041&.059&1.0145&-.5466&-.1758& .038&1.4632&-.7121&-.2346&.086\\
1998.1&1.3342&-.7732&-.2078&.065&1.0259&-.5679&-.1735& .039&1.4641&-.7090&-.2313&.088\\
1999.3&1.3342&-.7682&-.2097&.056&1.0204&-.5702&-.1636& .030&1.4642&-.7132&-.2424&.082\\
2000.8&1.3323&-.7654&-.2091&.060&1.0081&-.5077&-.2492& .064&1.4585&-.7086&-.2296&.068\\
2001.9&1.3342&-.7708&-.2014&.059&1.0196&-.5627&-.1823& .039&1.4636&-.7124&-.2426&.084\\
2003.1&1.3341&-.7691&-.2006&.057&1.0177&-.5686&-.1591& .029&1.4639&-.7144&-.2453&.082\\
2004.7&1.3341&-.7641&-.2102&.059&1.0094&-.5003&-.2642& .085&1.4635&-.7112&-.2453&.091\\
2006.0&1.3335&-.7705&-.2008&.058&1.0181&-.5543&-.1486& .038&1.4631&-.7136&-.2338&.094\\
2007.4&1.3335&-.7660&-.1982&.051&1.0149&-.5408&-.1174& .012&1.4645&-.7115&-.2378&.084\\
2008.7&1.3361&-.7700&-.2119&.076&1.0132&-.4941&-.1556& .045&1.4625&-.7112&-.2396&.081\\
2010.0&1.3334&-.7662&-.1988&.062&1.0230&-.4983&-.1529& .048&1.4623&-.7139&-.2405&.081\\
2010.9&1.3332&-.7665&-.1980&.064&1.0207&-.5140&-.1626& .058&1.4607&-.7150&-.2372&.077\\
2012.0&1.3324&-.7690&-.1950&.059&1.0332&-.5608&-.1197& .041&1.4616&-.7187&-.2424&.079\\
\enddata

\tablecomments{No fits are given for the 1992 session, as these data
  are very unreliable.  We do not recommend using these expressions
  for $nu>15$ GHz for dates prior to 1995, as the high frequency
  data for those dates have high errors.}

\end{deluxetable}

\subsection{3C123}

\begin{figure}[ht]
\centerline{{\includegraphics[width=2.4in]{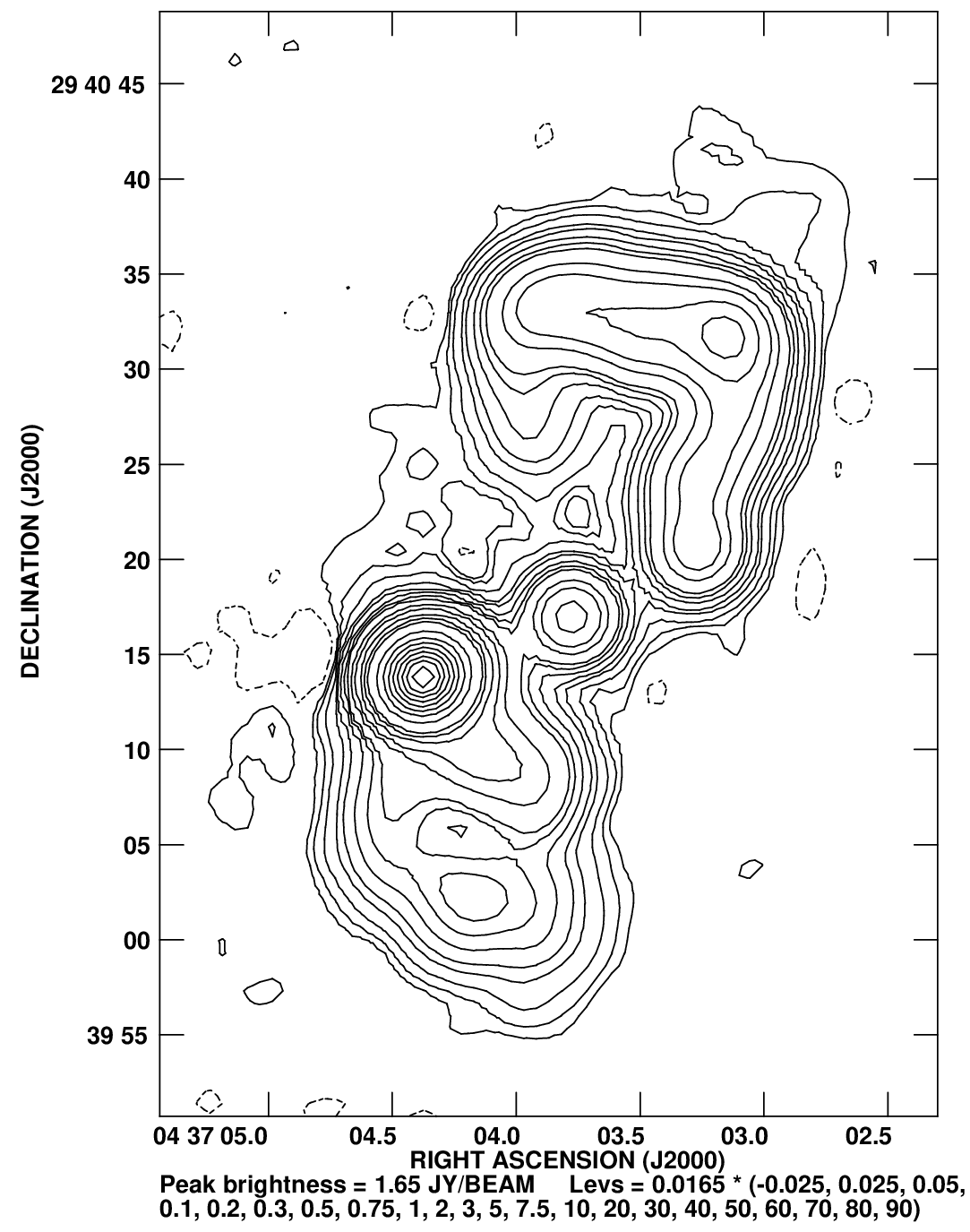}}
\hspace{3mm}{\includegraphics[width=3.7in]{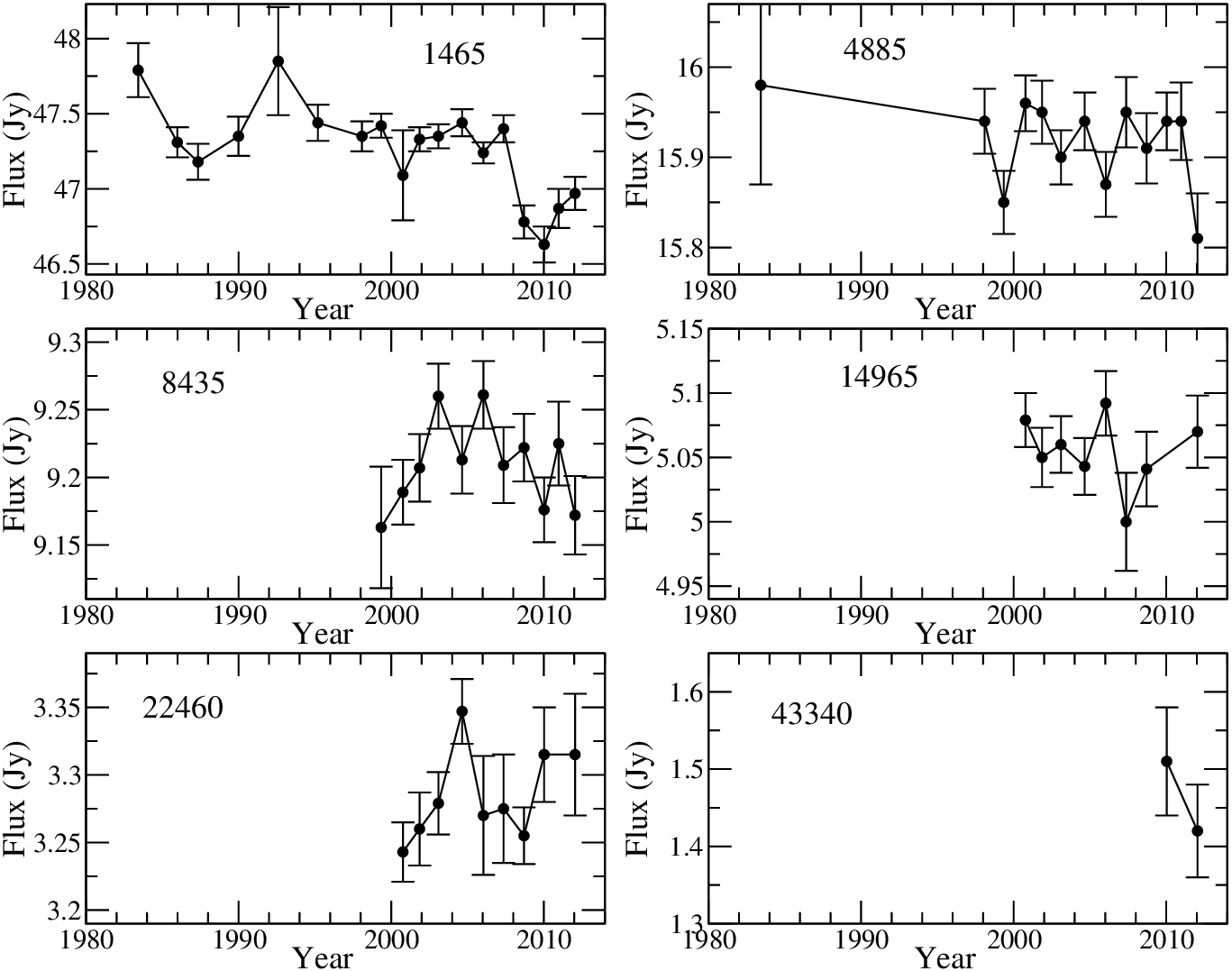}}}
\centerline{\parbox{6in}{
\caption{\small On the left panel, the structure of 3C123 at 22.485
  GHz, with 3 arcseconds resolution.  On the right panel, the secular
  variation of the observed flux density.}
\label{fig:3C123Maps}}}
\end{figure}

3C123 is a radio galaxy at redshift z=0.218.  With an angular size of
44 arcseconds, it is the largest of the objects in this study, and is
of limited use for the calibration of high-resolution interferometers.
We included it in this program primarily because it is a well-known
calibrator source at low frequencies.  The structure of the source,
with 3 arcseconds resolution at 22.5 GHz, utilizing the data taken in
this program, is shown in Fig.~\ref{fig:3C123Maps}.

3C123 is one of the non-variable objects that we have identified.  The
right panel of Fig.~\ref{fig:3C123Maps} shows the source's secular
flux density evolution, based on our polynomial expression for 3C286.
Notable is a small ($\sim$1\%) but significant drop at 1465 MHz,
beginning in 2008.  This drop is also seen at 1275 MHz in 2008, but at
no other frequency utilized between 2008 and 2011.  We believe the
drop is real, but have no explanation for its origin.

\subsection{3C138}

\begin{figure}[ht]
\centerline{{\includegraphics[height=2.4in,angle=-90,origin=c]{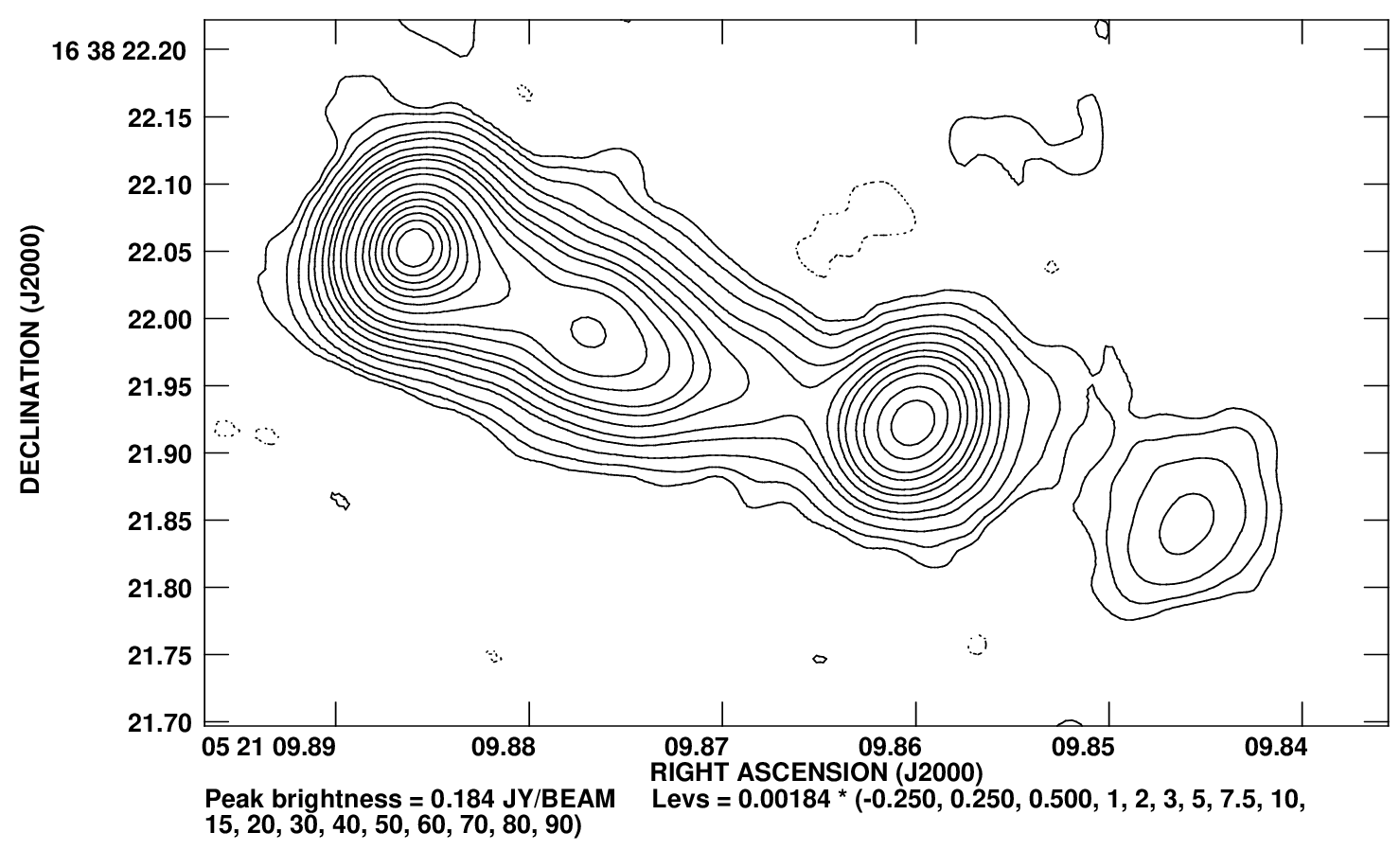}}
\hspace{3mm}{\includegraphics[width=3.7in]{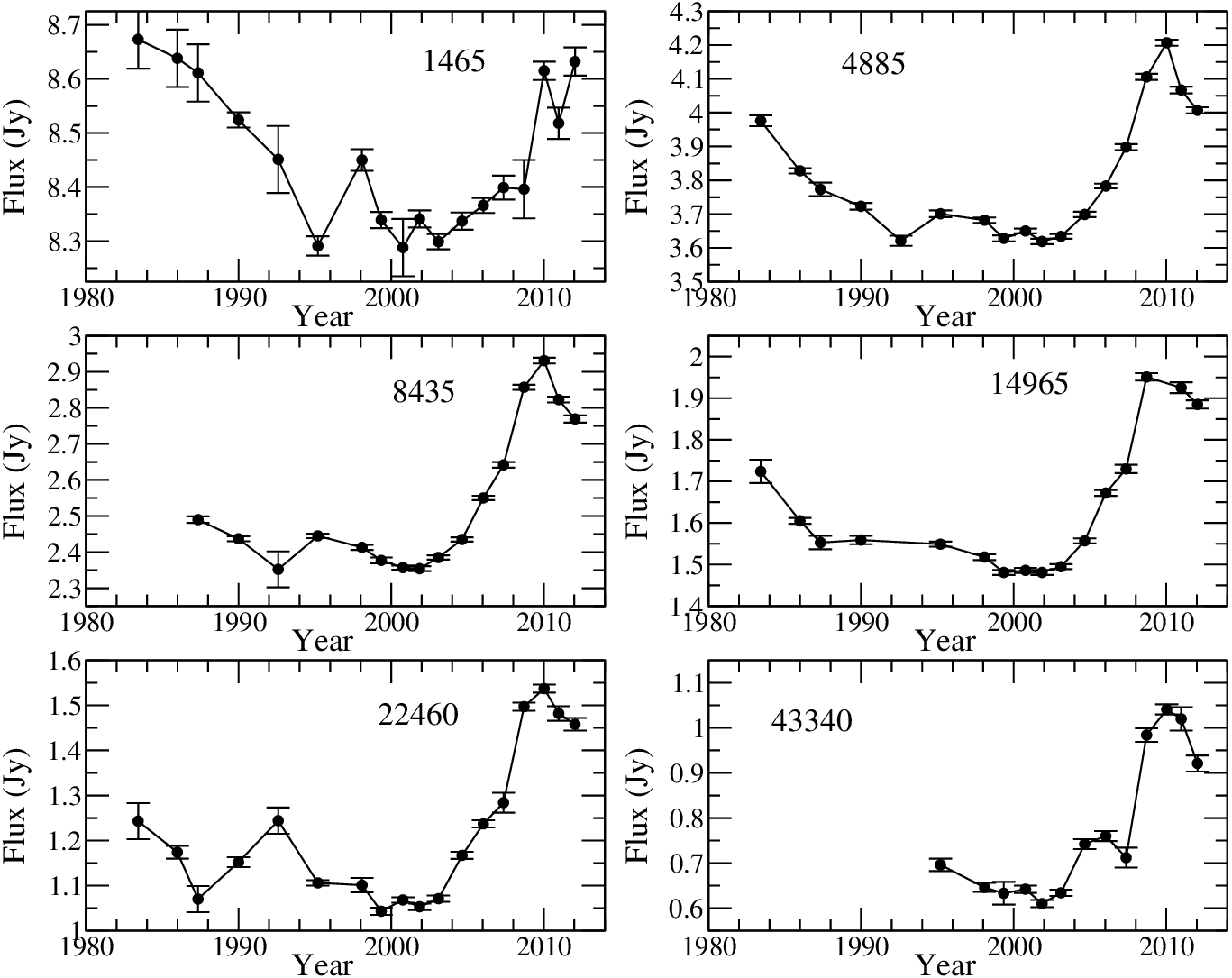}}}
\centerline{\parbox{6in}{
\caption{\small On the left panel, the structure of 3C138 at 43 
  GHz, with 60 milliarcseconds resolution.  On the right panel, the secular
  variation of the observed flux density.}
\label{fig:3C138Maps}}}
\end{figure}

3C138 is a compact steep-spectrum quasar at redshift z=0.759.  VLBI
images show it has structure on a maximum scale of 0.62 arcseconds,
with a highly polarized, one-sided jet leading from the nucleus, and
some weak structure on the other side (\citet{Cot97a}).  These
structures are shown in a 60 milliarcsecond resolution VLA image at 43
GHz shown in Fig.~\ref{fig:3C138Maps}.  Due to its high flux density
and small angular size, this source is commonly utilized by the VLA
for amplitude gain calibration.

3C138 is not one of the \citet{Baa77} sources, probably because it was
known to be significantly variable.  Our data show that this is indeed
the case, with the slow decline noted after 1983 abruptly terminated
with a large increase beginning in 2003, as shown in the right-hand
side of Fig.~\ref{fig:3C138Maps}.  The flare, with the proportional
increase being greatest at the higher frequencies (nearly doubling the
flux density at 43 GHz), reached a peak in 2010, and is now quickly
subsiding.  We have fitted this variable spectrum with a cubic
polynomial function for each of the 18 session, whose coefficients are
given in Table~\ref{tab:CalTimeSpec}.

\subsection{3C147}

\begin{figure}[ht]
\centerline{{\includegraphics[height=2.7in]{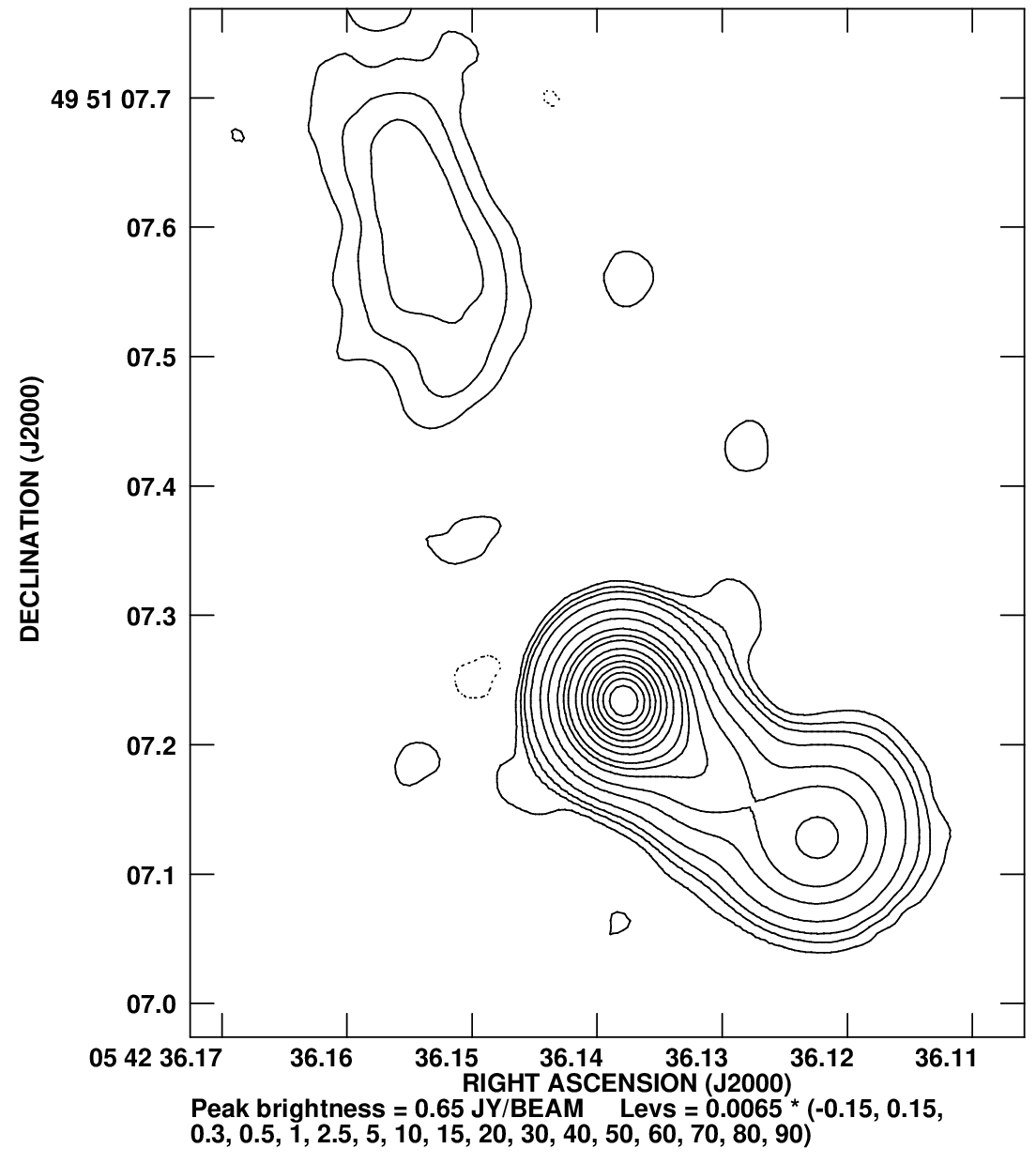}}
\hspace{3mm}{\includegraphics[width=3.6in]{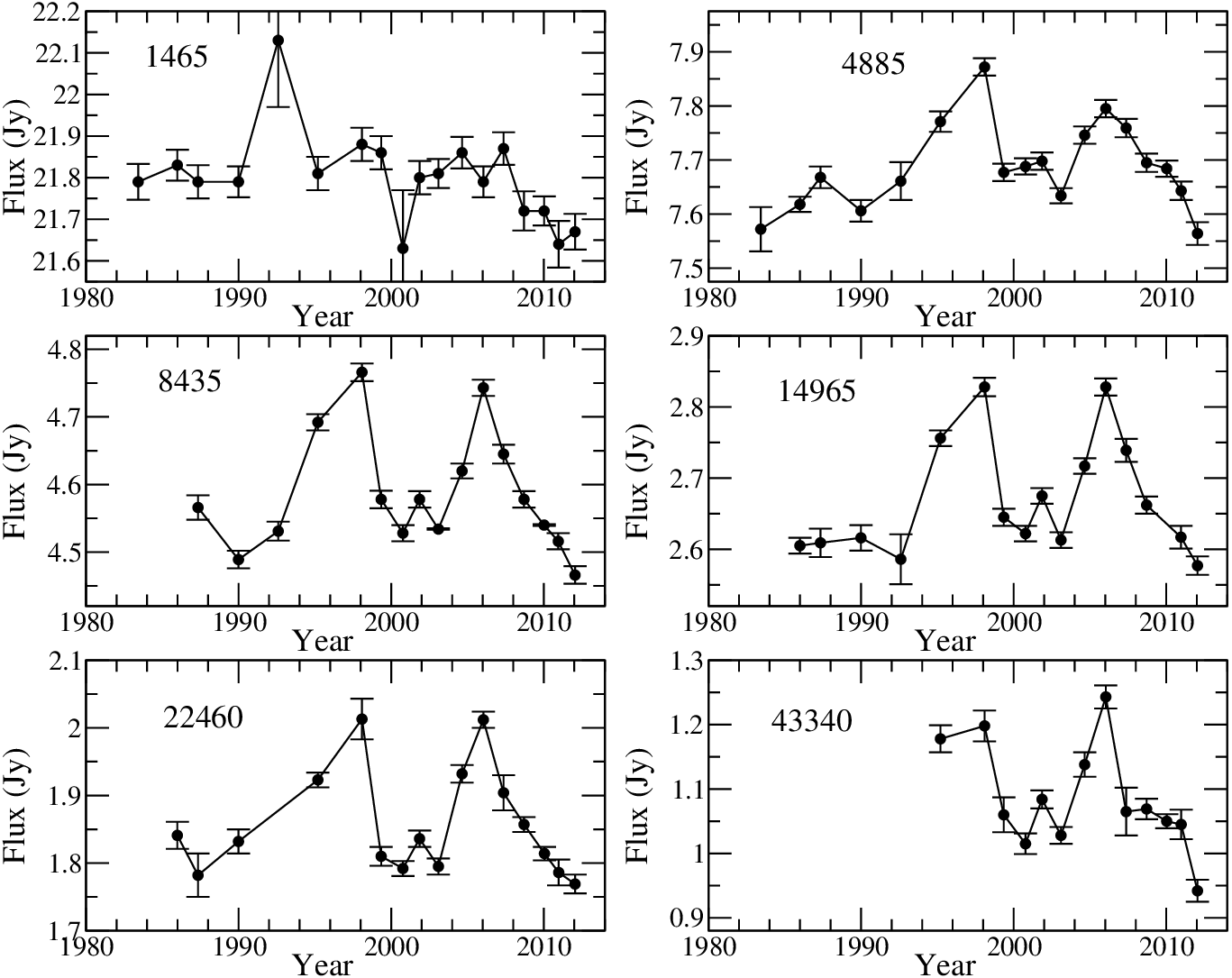}}}
\centerline{\parbox{6in}{
\caption{\small On the left panel, the structure of 3C147 at 43 
  GHz, with 60 milliarcseconds resolution.  On the right panel, the secular
  variation of the observed flux density.}
\label{fig:3C147Maps}}}
\end{figure}

3C147 is a compact steep-spectrum quasar at redshift z = 0.545.  VLBI
imaging shows structure of maximum extent of 200 milliarcseconds
(\citet{Ross09}).  A VLA high resolution image at 43 GHz, made with
high resolution data, is shown in the left panel of
Fig.~\ref{fig:3C147Maps}.  The small angular size and high spectral
flux density of 3C147 have resulted in it being commonly used as a
flux density scale calibrator by the VLA.
This source has undergone significant changes in flux density over
time, as shown in the right panel of Fig.~\ref{fig:3C147Maps}.  For
all bands other than 20cm, a rise of 5 to 10\% was seen between 1990
and 1998, again with the larger increases being seen at higher
frequencies.  This increase was followed by a sudden drop of similar
magnitude lasting until 2004, then another short and rapid rise, of
similar magnitude, to 2006.  Since then, the flux density has been
dropping steadily.  We have fitted the variable flux densities of this
source with cubic polynomials for each of the 18 observing session,
the results of which are shown in Table~\ref{tab:CalTimeSpec}.

\subsection{3C196}

\begin{figure}[ht]
\centerline{\hbox{\includegraphics[width=2.6in]{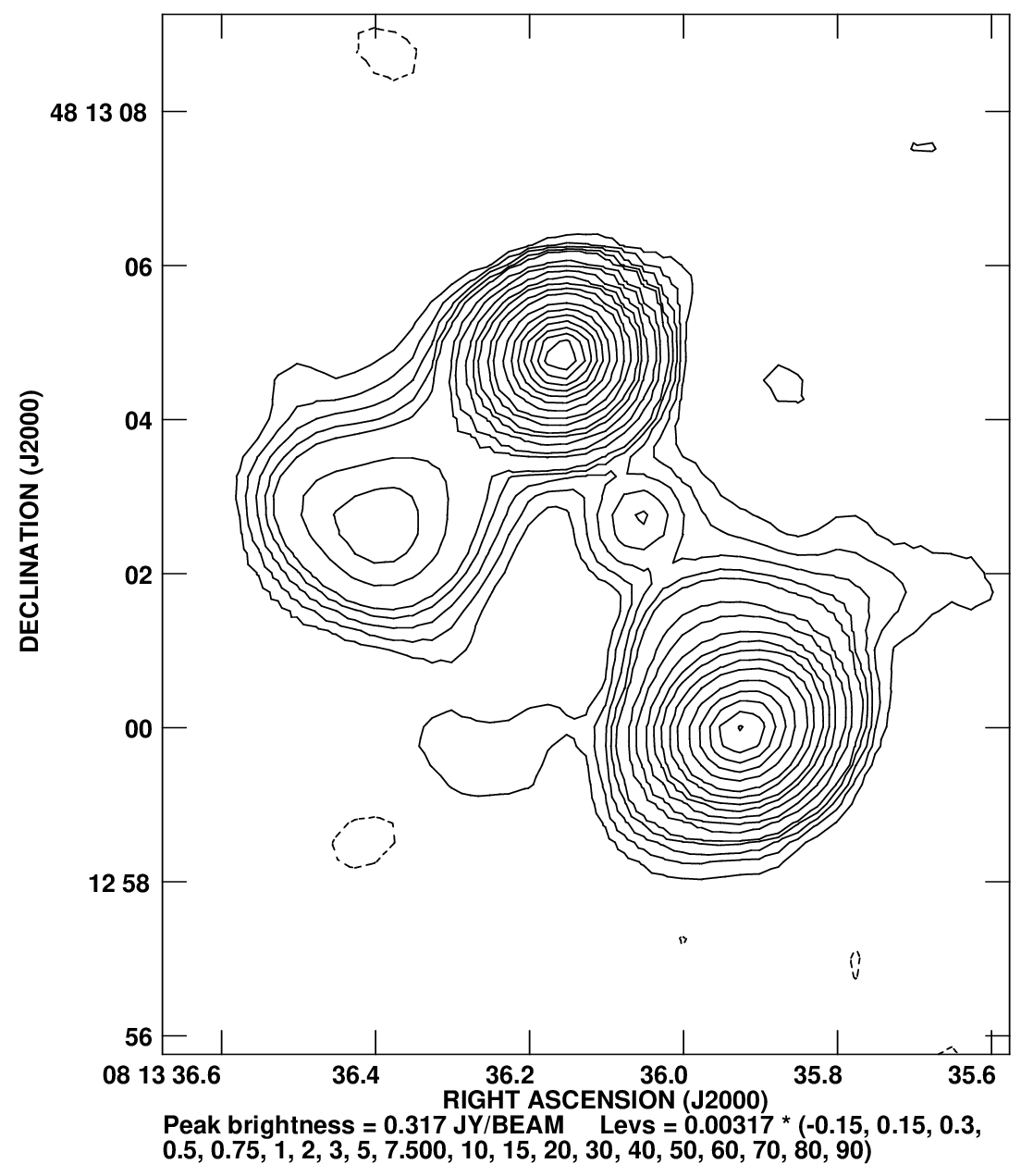}}
\hspace{5mm}
\hbox{\includegraphics[width=3.7in]{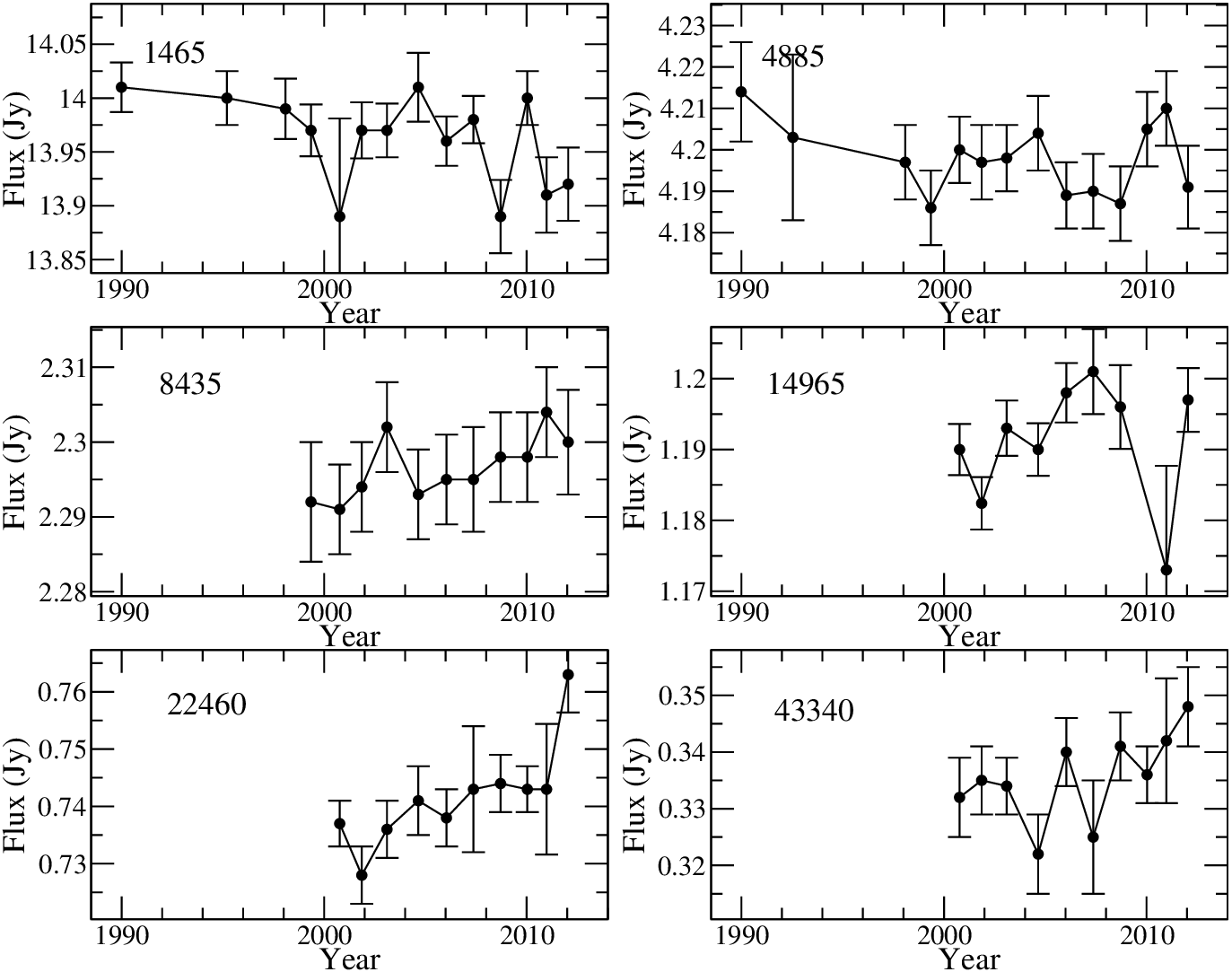}}}
\centerline{\parbox{6in}{
\caption{\small Left panel: The structure of 3C196 at 22 GHz, with 1
  arcsecond resolution.  Right panel:  The observed secular variation
  of this source.}
\label{fig:3C196Maps}}}
\end{figure}

3C196 is a quasar at redshift z = 0.871.  The source comprises a pair
of compact lobes symmetrically placed about a weak core, with maximum
angular extent of about 7 arcseconds, as shown in the 22 GHz, 1
arcsecond resolution image in the left panel of
Fig.~\ref{fig:3C196Maps}, made from the data taken for this program.
In the right panel is shown the secular variation.
This source is one of the non-variable objects we have identified.
Because of this, its small angular size, and remarkably straight
spectrum (\citet{Sca12}), 3C196 is an excellent flux density
calibration source for low-resolution observations for frequencies up
to $\sim$10 GHz.

\subsection{3C286}

\begin{figure}[ht]
\centerline{\hbox{\includegraphics[height=3.1in,angle=-90]{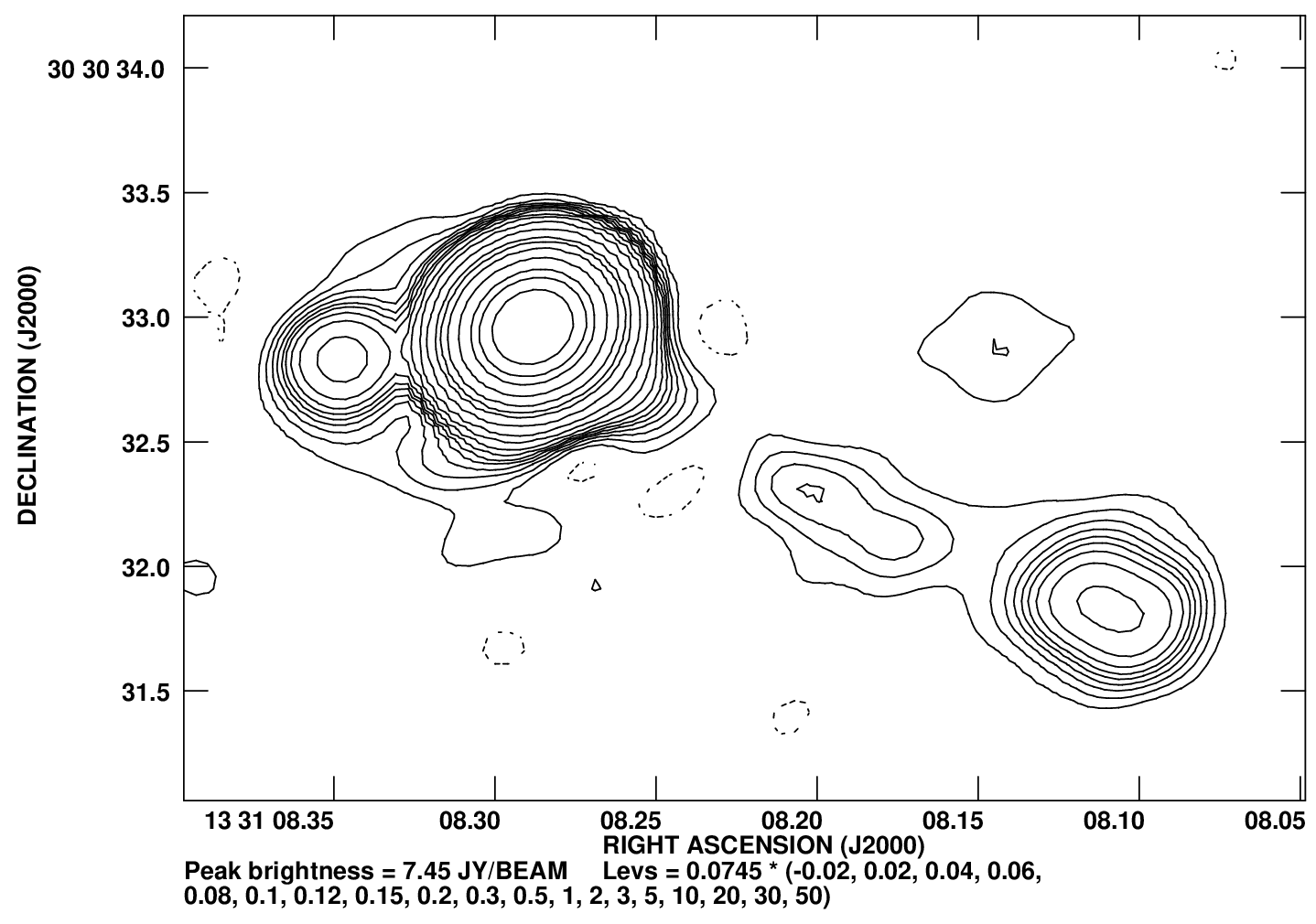}}
\hspace{3mm}{\includegraphics[height=3.1in,angle=-90]{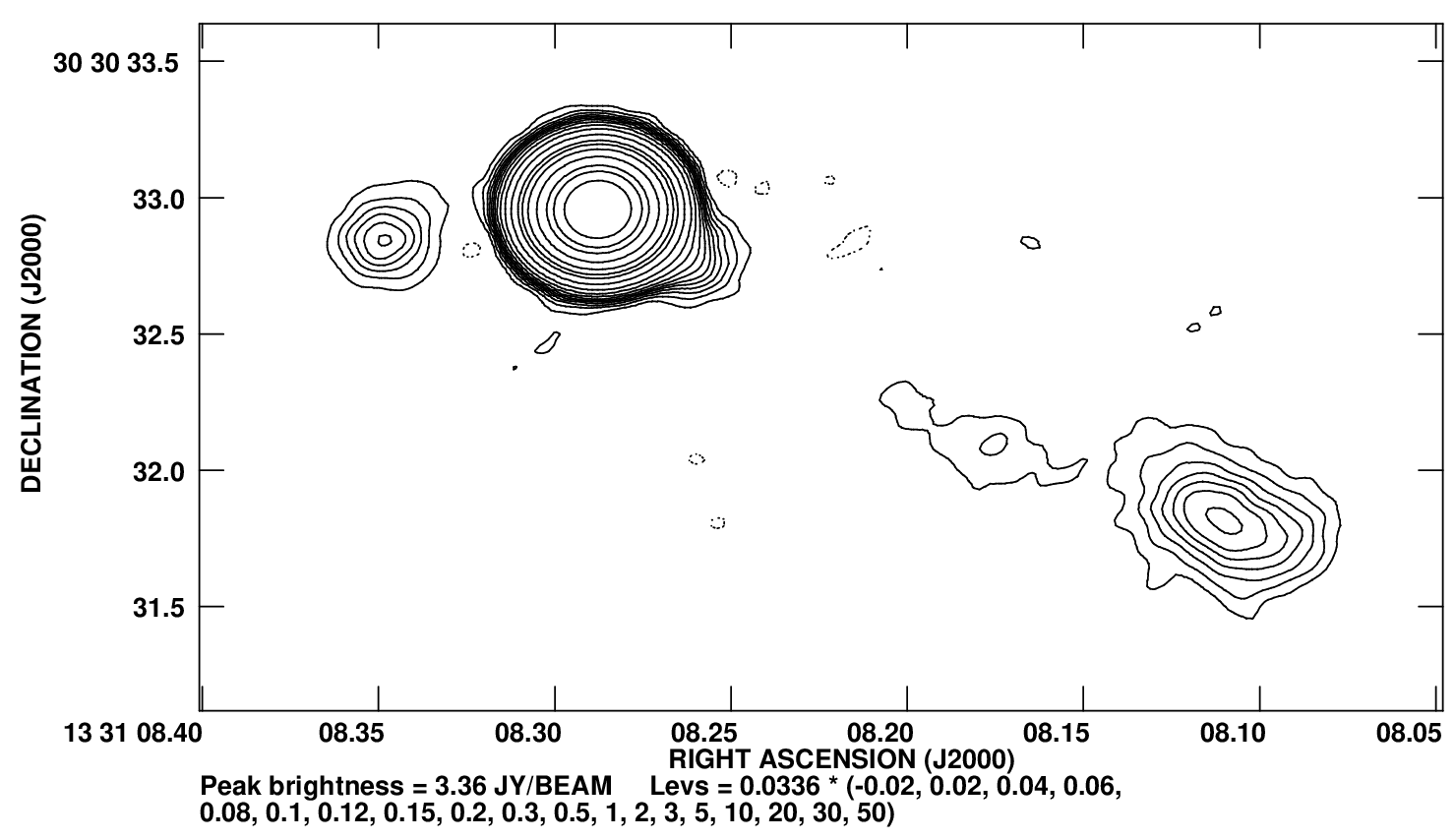}}
}
\centerline{\parbox{6in}{
\caption{\small The structure in 3C286 at 5 GHz (left) with 310
  milliarcsecond resolution, and at 15 GHz (right), with 220
  milliarcsecond resolution.}
\label{fig:3C286Maps}}}
\end{figure}

3C286 is a compact steep-spectrum quasar at redshift z=0.846.  High
resolution VLA images show the source has a steep-spectrum extension
of length $\sim$2.5 arcseconds to the SW, and a smaller knot of
emission to the east, as shown in Fig.~\ref{fig:3C286Maps}.  VLBI,
EVN, and MERLIN images show a highly polarized linear structure of
maximum 50 milliarcseconds extent (\citet{Cot97b}, \citet{Jia96},
\citet{Aku95}).
The milliarcsecond structure of this source has some very unusual
features.  The VLBA imaging shows that the compact central structure
shown in the figure is steep spectrum, resolved to the VLBA, and
highly polarized throughout.  There appears to be no optically thick
central nuclear emission.  It is doubtless the lack of the nucleus,
and the uniform polarization visible in the maps of \citet{Cot97b}
that make this object an extraordinarily stable and useful calibrator.

\subsection{3C295}

\begin{figure}[ht]
\centerline{\hbox{\includegraphics[width=2.8in]{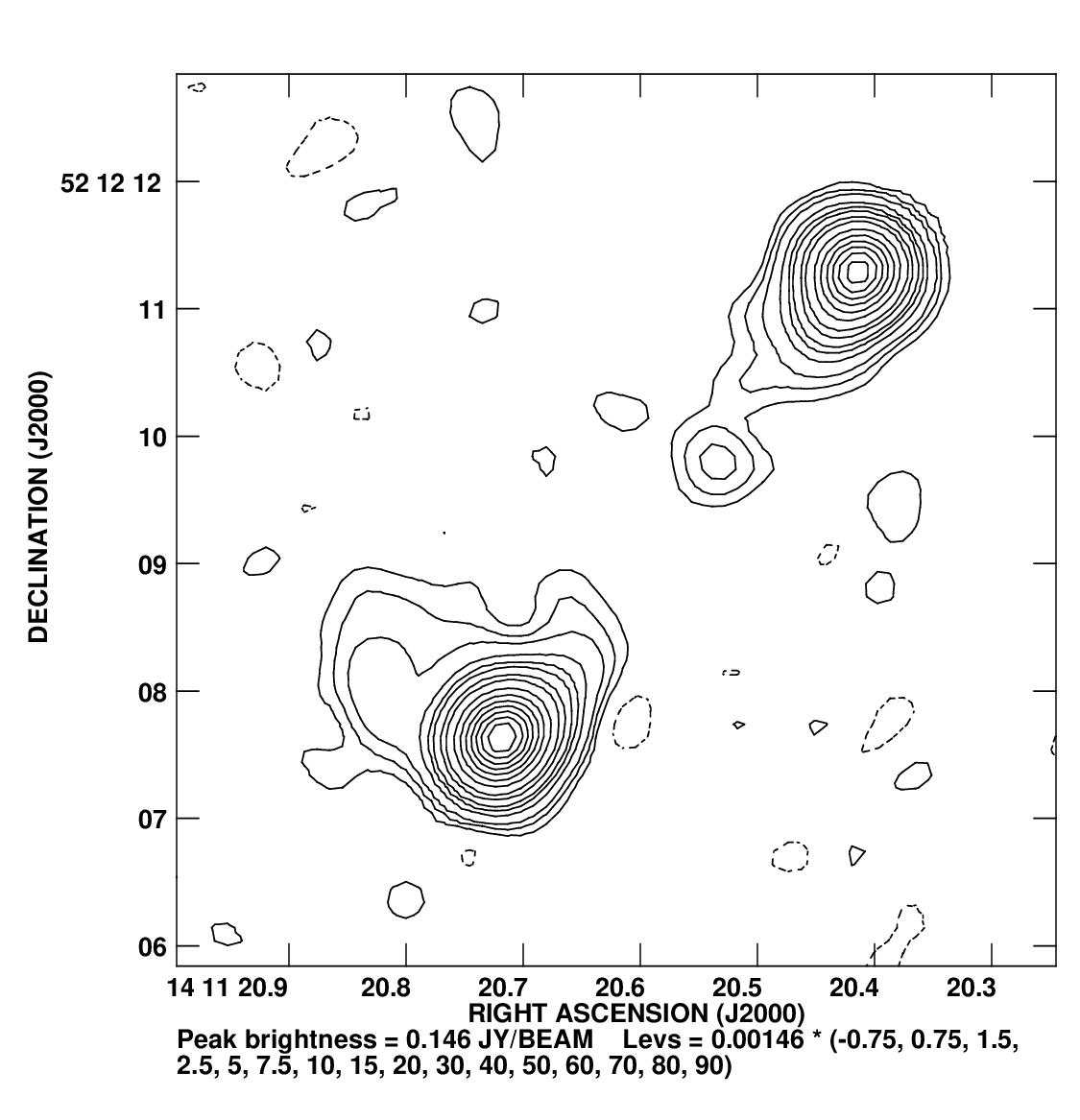}}
\hspace{5mm}
\hbox{\includegraphics[width=3.5in]{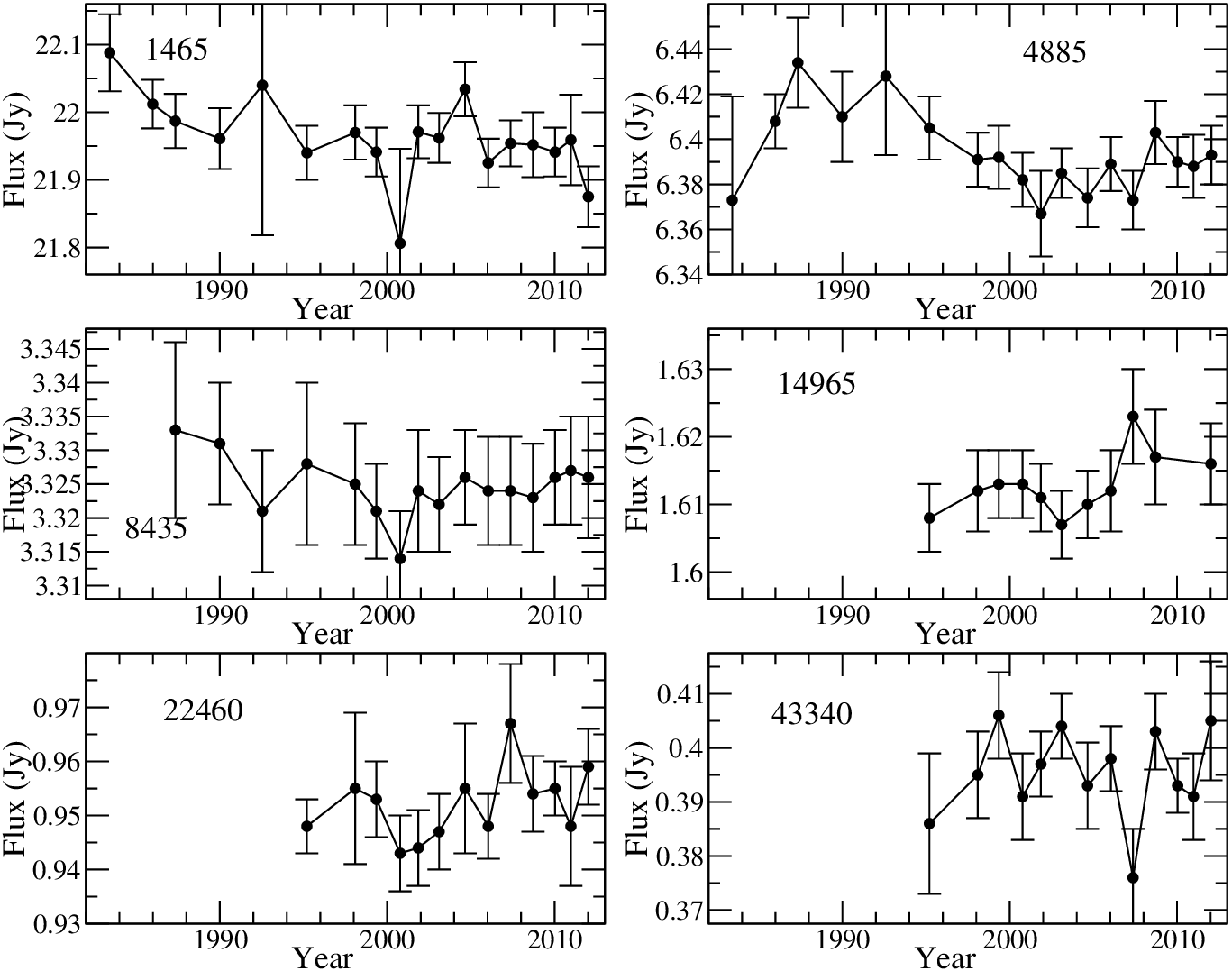}}}
\centerline{\parbox{6in}{
\caption{\small (left) The structure of 3C295 at 43 GHz with 1
  arcsecond resolution.  (right) The secular variation of the flux
  density in 3C295 -- the source is clearly non-variable.}
\label{fig:3C295Maps}}}
\end{figure}

3C295 is a radio galaxy at redshift z = 0.464.  This source is a small
double of $\sim$5 arcseconds extent with very weak central nucleus,
contributing less than 1\% of the total flux density at 15 GHz
(\citet{Tay92}).  An image of the source at 23 GHz with 1 arcsecond
resolution, utilizing our data, is shown in the left panel of
Fig.~\ref{fig:3C295Maps}. In the right panel is shown the secular
variation of the spectral flux density.  

The weakness of the nuclear core emission suggests that this source
should have a stable flux density, as noted by \citet{Ott94}.  Our
observations confirm this suggestion, showing no variation in its flux
density to a level below 1\% over the duration of our program.  3C295
is an excellent flux density calibration source for low-resolution
interferometers below $\sim$ 10 GHz.

\subsection{NGC6572}

\begin{figure}[ht]
\centerline{\hbox{\includegraphics[width=2.6in]{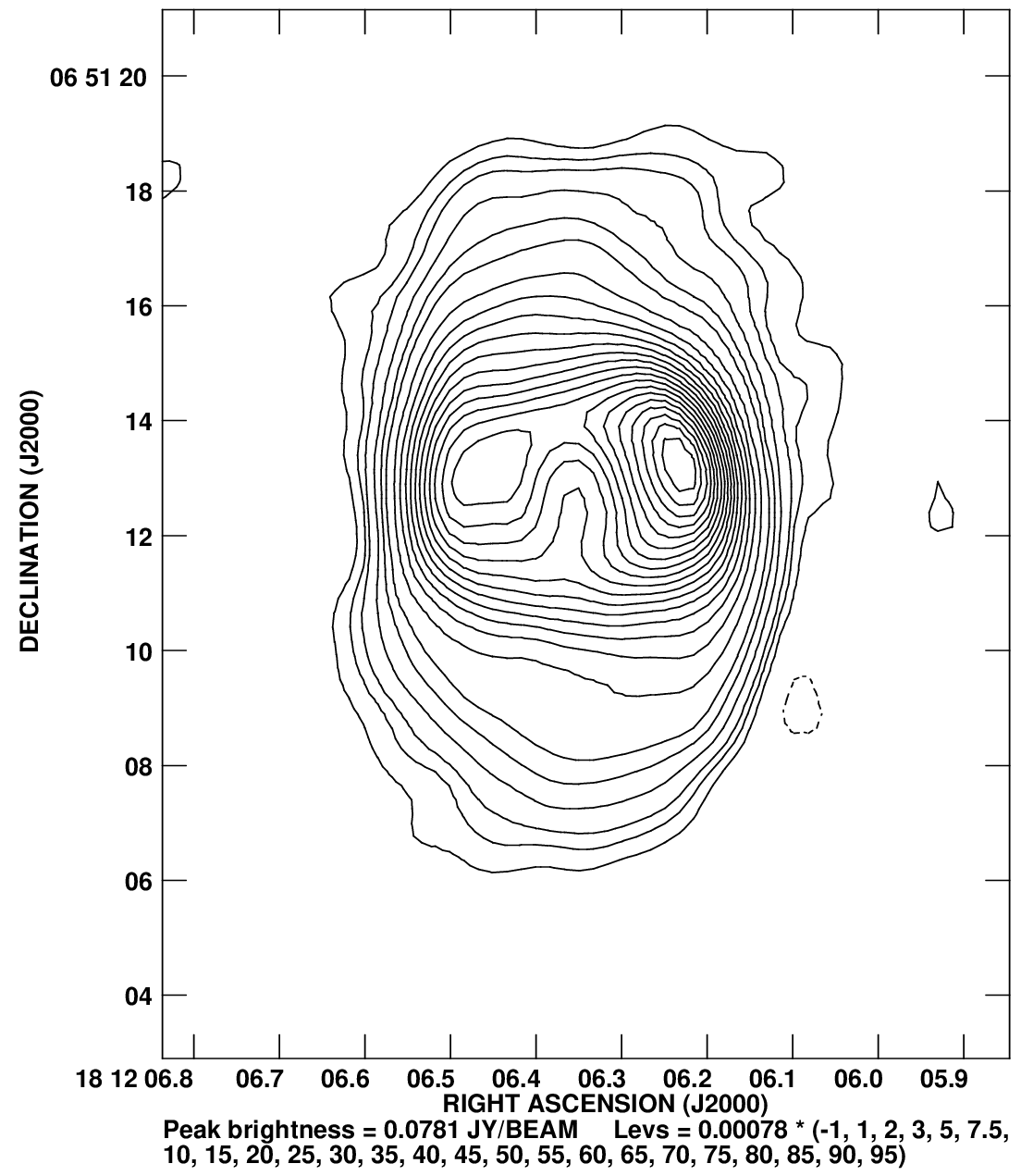}}
\hspace{5mm}
\hbox{\includegraphics[width=3.6in]{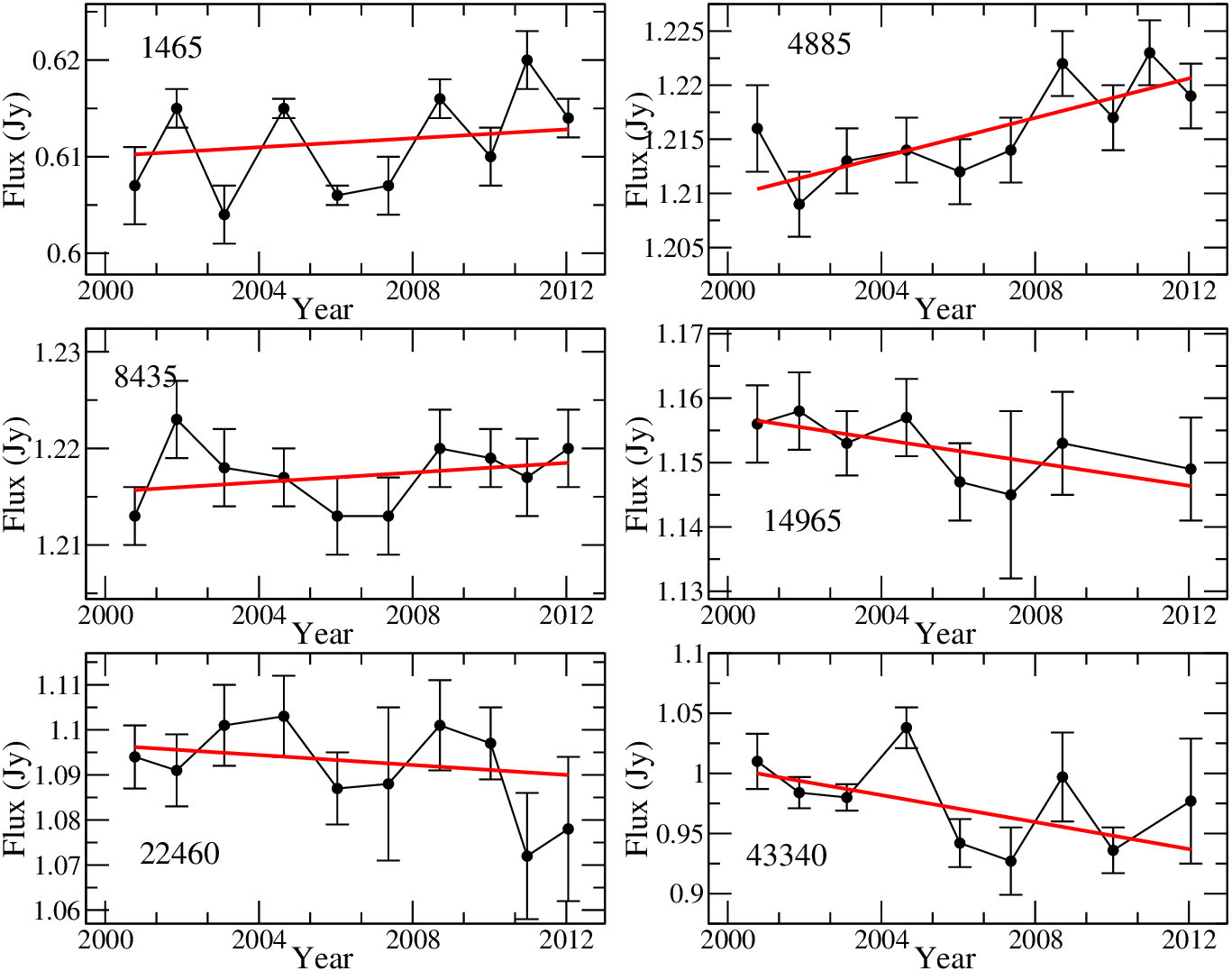}}}
\centerline{\parbox{6in}{
\caption{\small The structure of NGC6572:  (left) At 22 GHz with 1.1
  arcsecond resolution; (right) The secular variation in the flux
  density.  There is marginal evidence for a secular evolution of the
  flux density.  The results of a weighted linear fit to the data are
  superposed.}
\label{fig:N6572Image}}}
\end{figure}

The planetary nebula NGC6572 was added to this program in 2000, based
on its apparent similarity to NGC7027. The source is about 13
arcseconds in angular extent, with very diffuse emission extended to
the north and south, as shown in Fig.~\ref{fig:N6572Image}.  The lack
of sharp edge emission makes the source a poor calibrator, as longer
interferometer spacings provide insufficient amplitude to permit a
stable gain solution for antennas located at the ends of the array.
There is a possible trend in this source's flux density over a 12-year
timescale, as shown in the right-hand panel of the figure, showing the
data, and a linear fit.
The results of the fits are shown in Table~\ref{tab:ChangePN}.
\begin{deluxetable}{ccccccccccc}
\tabletypesize{\scriptsize} 
\tablecaption{Secular Evolution of NGC6572 and NGC7027 
\label{tab:ChangePN}}
\tablewidth{0pt} 
\tablehead{ \colhead{Freq}&\multicolumn{4}{c}{NGC6572}&
\multicolumn{6}{c}{NGC7027}\\ 
&\colhead{S2000} &\colhead{Error}&\colhead{Slope}& \colhead{Error}&
 \colhead{S2000} &\colhead{Error}&\colhead{Slope}& \colhead{Error}&
 \colhead{Accel.}&\colhead{Error}\\
\colhead{MHz}&\colhead{mJy}&\colhead{mJy}&\colhead{mJy/Yr}&\colhead{mJy/Yr}&
              \colhead{mJy}&\colhead{mJy}&\colhead{mJy/Yr}&\colhead{mJy/Yr}&
              \colhead{mJy/$yr^2$}&\colhead{mJy/$yr^2$}} 
\startdata 
1465 & 610& 1&0.23&0.20&1561&2&3.6&0.2&-.05&0.02\\
4885 &1210& 2&0.91&0.27&5392&5&-3.0&0.4&0.12&0.05\\
8435 &1216& 2&0.25&0.30&5851&5&-7.1&0.7&&\\
14965&1157& 4&-0.9&0.7 &5702&7&-7.4&1.1&&\\
22460&1097& 5&-0.6&0.8 &5480&13&-7.0&2.0&&\\
43340&1004&11&-5.6&2.1 &5091&42&-5.0&6.3&&\\
\enddata 
\end{deluxetable}

\subsection{NGC7027}

\begin{figure}[ht]
\centerline{\hbox{\includegraphics[width=2.8in]{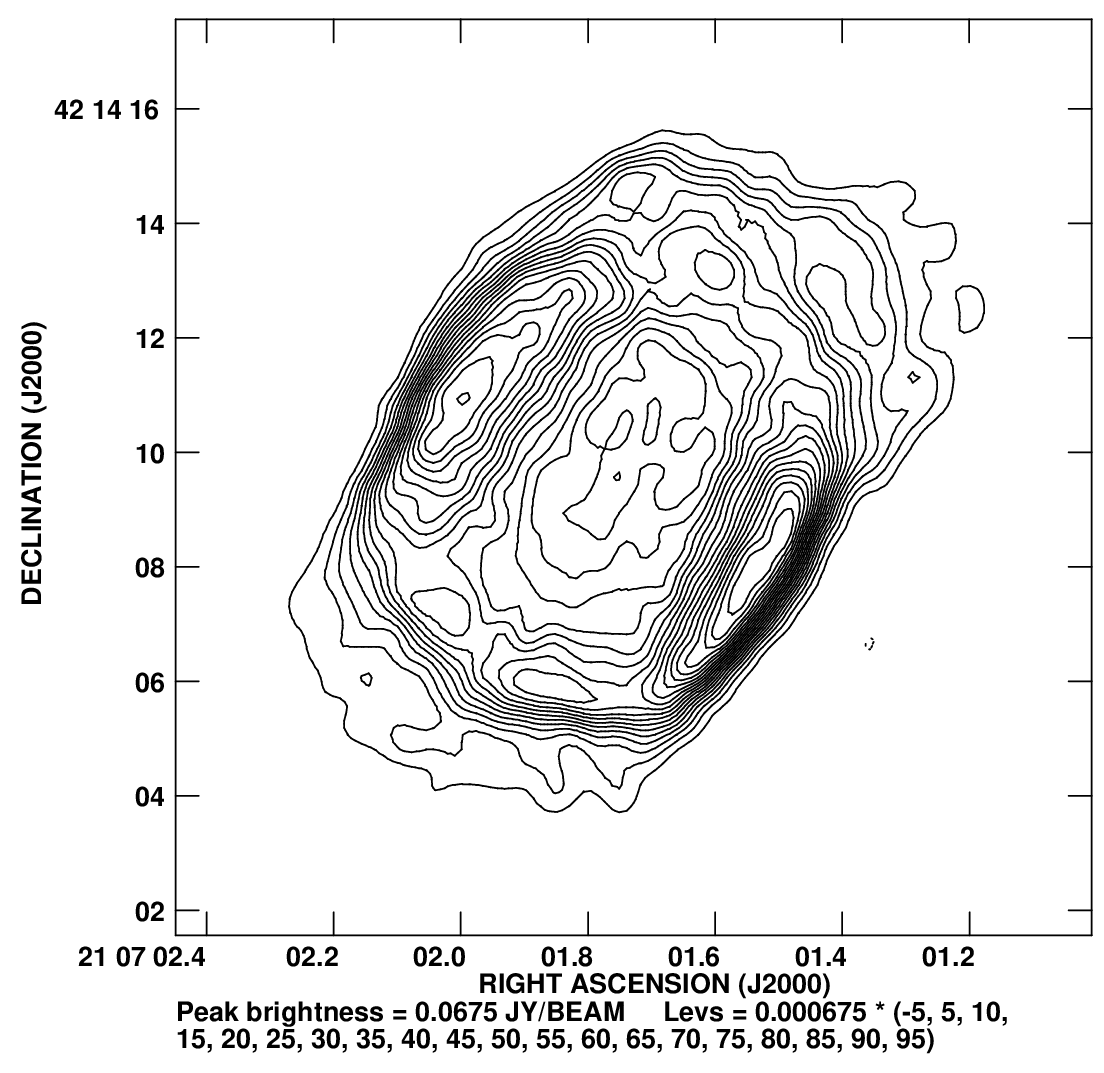}}
\hspace{3mm}
\hbox{\includegraphics[width=3.6in]{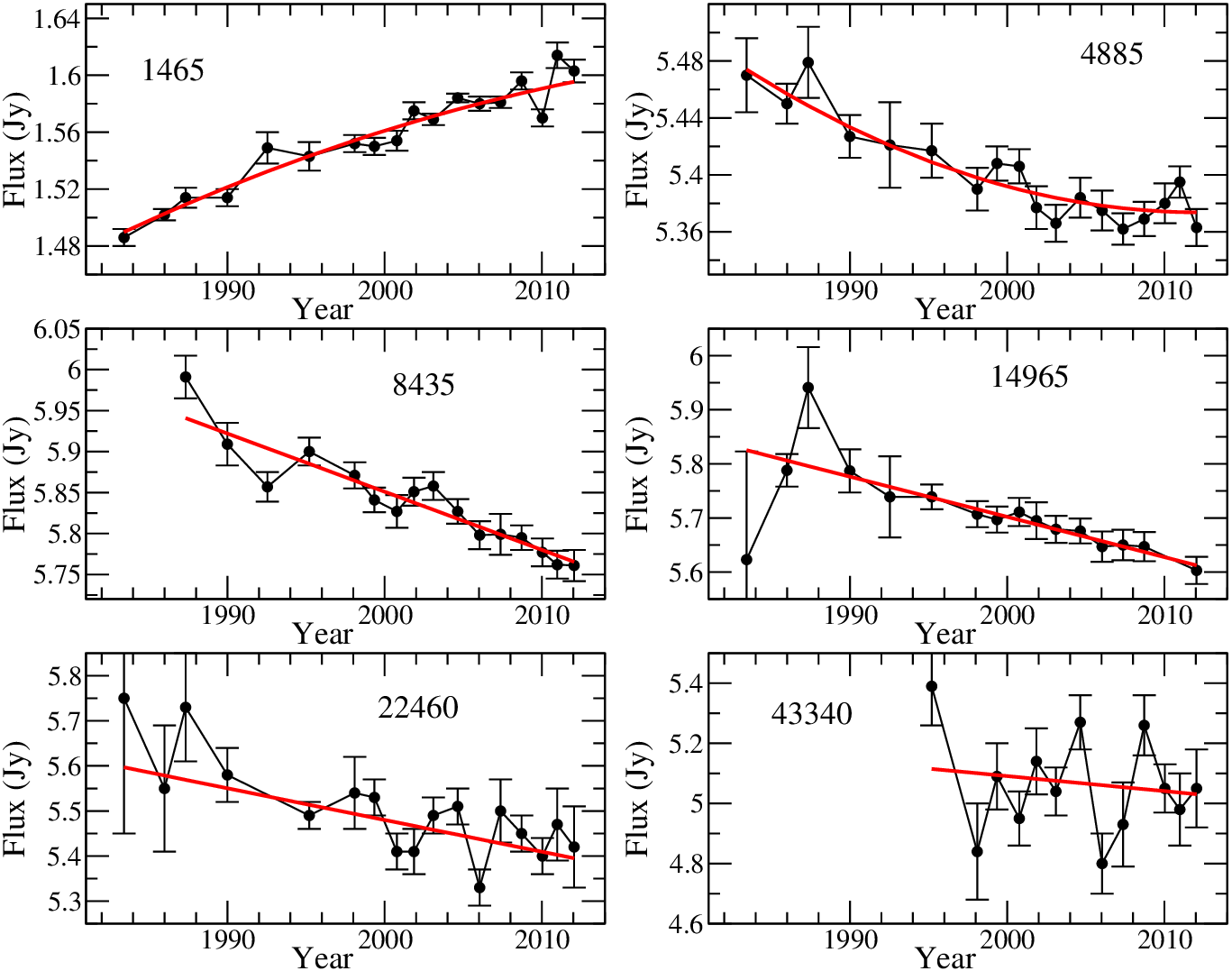}}}
\centerline{\parbox{6in}{
\caption{\small (left) The structure of NGC7027 at 43 GHz with 600
milliarcsecond resolution; (right) The secular variation of the source
over the 30-year span of this program. The flux density is decreasing
at all frequencies where the source is optically thin, and is
increasing at 1465 MHz, where the source is optically thick.  There is
marginal evidence for a deceleration at the lowest two frequencies.}
\label{fig:N7027Maps}}}
\end{figure}

The bright, nearby planetary nebula NGC7027 is one of the
\citet{Baa77} sources. An image of the source at 43 GHz, made with our
data and at 0.6 arcseconds resolution is shown in
Fig.~\ref{fig:N7027Maps}.  Because of its large angular size, this
source is not a good primary calibrator for high-resolution
interferometers.
NGC7027 has long been known to be increasing in flux density at the
frequencies where it is optically thick, and to be decreasing at
frequencies where it is optically thin.  \citet{Zijl08}, using the
data taken in this program up to 2006, utilized these secular
variations, combined with optical observations of the linear expansion
velocity and photo-ionization models, to place tight limits on the age
and distance to the planetary nebula, and on the mass of the central
star.  The observations taken since 2006 show the flux density
continuing to change as expected, but there is now some evidence of a
deceleration in the secular increase at 1465 MHz, and also a possible
deceleration in the secular decrease at 4885 MHz.  The results of a
weighted fit are given in in Table~\ref{tab:ChangePN}, and plotted in
the figure.  We emphasize that the statistical significance for
deceleration is low -- further long-term monitoring will be needed to
establish the reality of this result.  

\subsection{MWC349}

\begin{figure}[ht]
\centerline{\hbox{\includegraphics[width=2.7in]{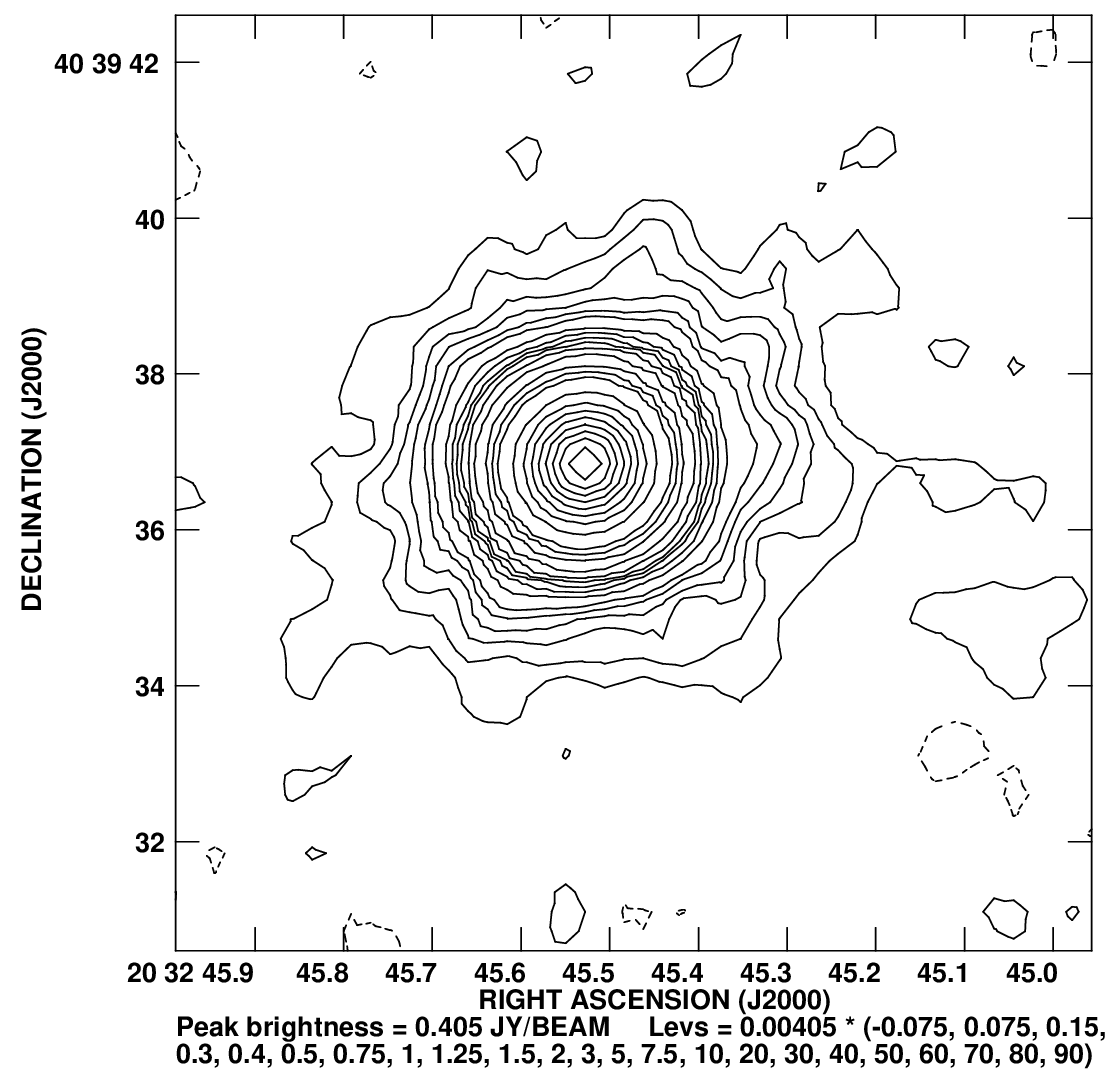}}
\hspace{3mm}
\hbox{\includegraphics[width=3.6in]{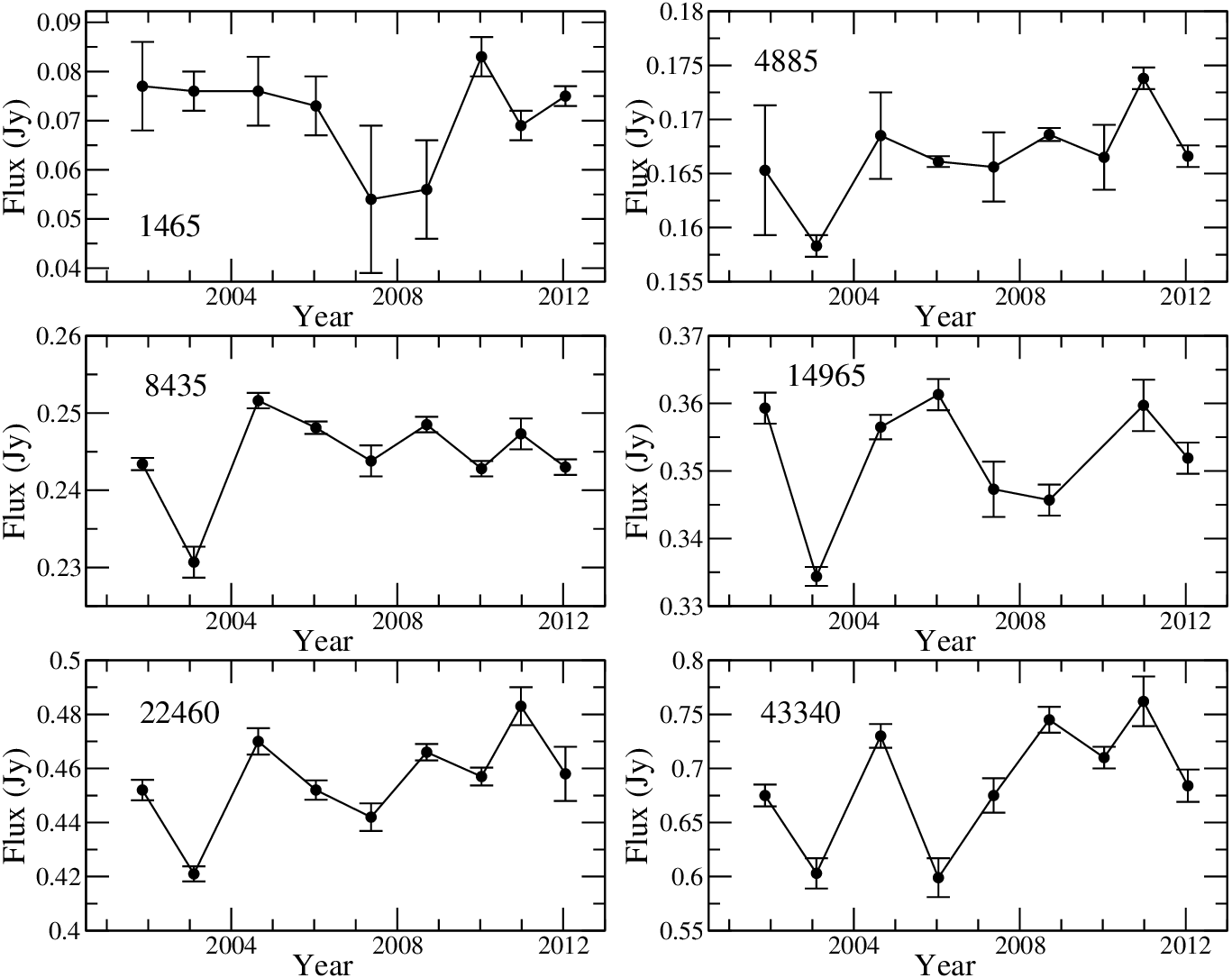}}}
\centerline{\parbox{6in}{
\caption{\small (Left) The structure of MWC349 at 22 GHz with 1.1
  arcseconds resolution.  A diffuse region of about 8 arcseconds
  extent surrounds the central core.  (Right) Temporal Changes in the
  flux density for MWC349.  The source shows significant short-term
  (one year, or less) variations in flux density.  }
\label{fig:MWC349}}}
\end{figure}

MWC349 is a binary Be star of 2.4 arcseconds separation. We added this
source to our monitoring list, following suggestions that it may be
suitable as a flux density calibrator at long-millimeter and
centimeter wavelengths. Its radio structure has been extensively
studied by \citet{Taf04} showing 0.3 arcsecond extent at 10 GHz, and 1
arcsecond at 3 GHz. A new, deeper image at 23 GHz with 1.1 arcseconds
resolution with our data is shown in Fig.~\ref{fig:MWC349}.  The
source is known to be variable in the visible, IR, and mm wavelengths
(\citet{Taf04}, and references therein).

Our observations show there are peculiar changes in the flux density
of significant magnitude seen simulataneously over most bands.  The
most notable was in the Feb 2003 observations, when a drop in flux
density of about 10\% was seen at all frequencies except 1465 MHz.
This drop cannot be due to pointing errors or atmospheric absorption,
since these always affect high frequency observations much more
strongly.  A similar, but smaller, rise was noted in the Dec 2010
observations.  It would appear that MWC349 has small but significant
($\sim$10\%) changes in flux density.  The changes are of short
duration -- less than one year.  Because of these, the source is not a
suitable absolute flux density calibrator for cm or mm wavelengths.

\section{The Planets}

\begin{figure}[ht]
\centerline{\hbox{
\includegraphics[width=5in]{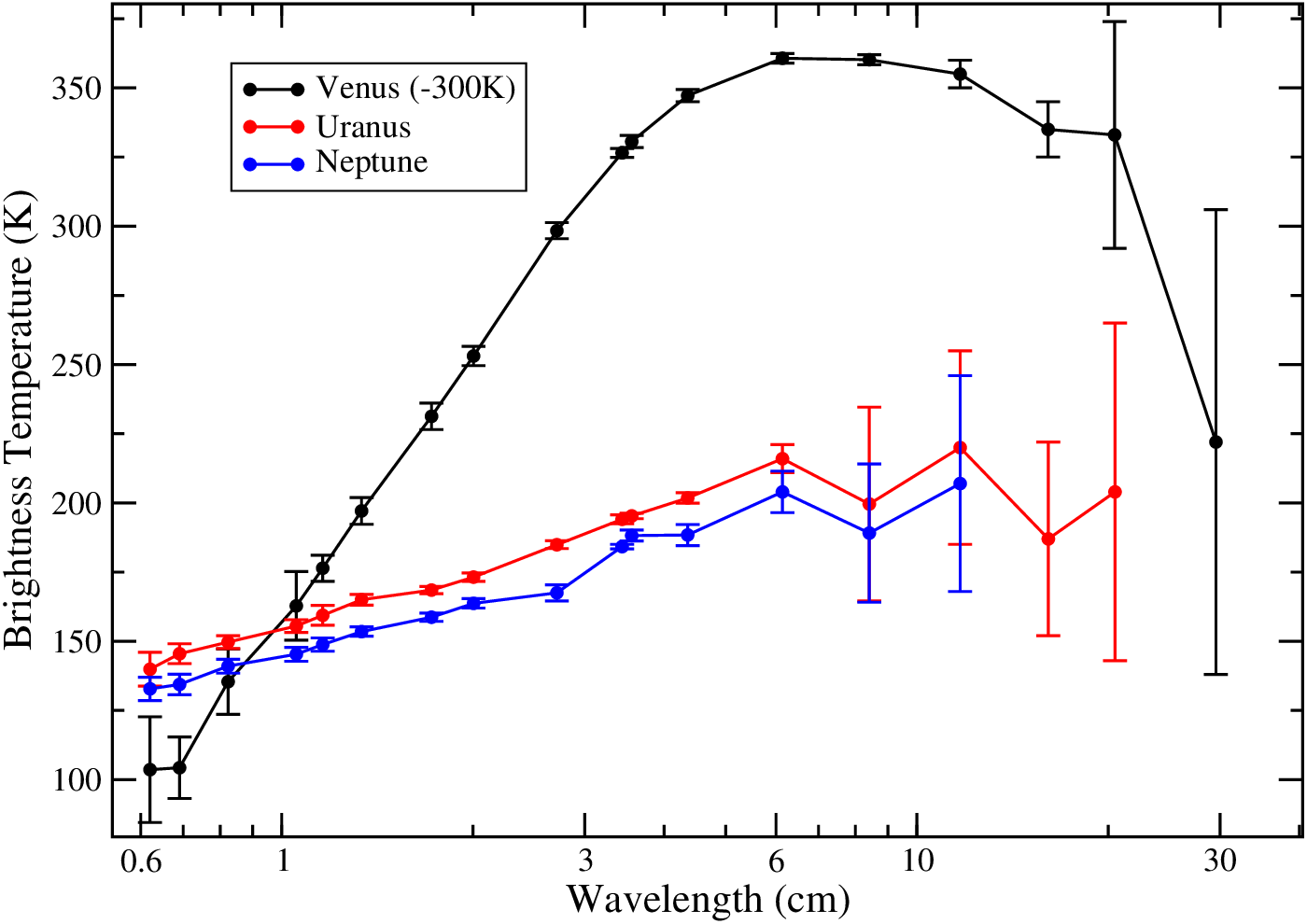}}}
\centerline{\parbox{6in}{
\caption{\small The brightness spectrum of the planets Venus, Uranus,
  and Neptune in 2012. Note that the brightness of Venus has been
  reduced by 300K.  }
\label{fig:PlanetSpectrum}}}
\end{figure}

\begin{figure}[ht]
\centerline{\hbox{
\includegraphics[width=2in]{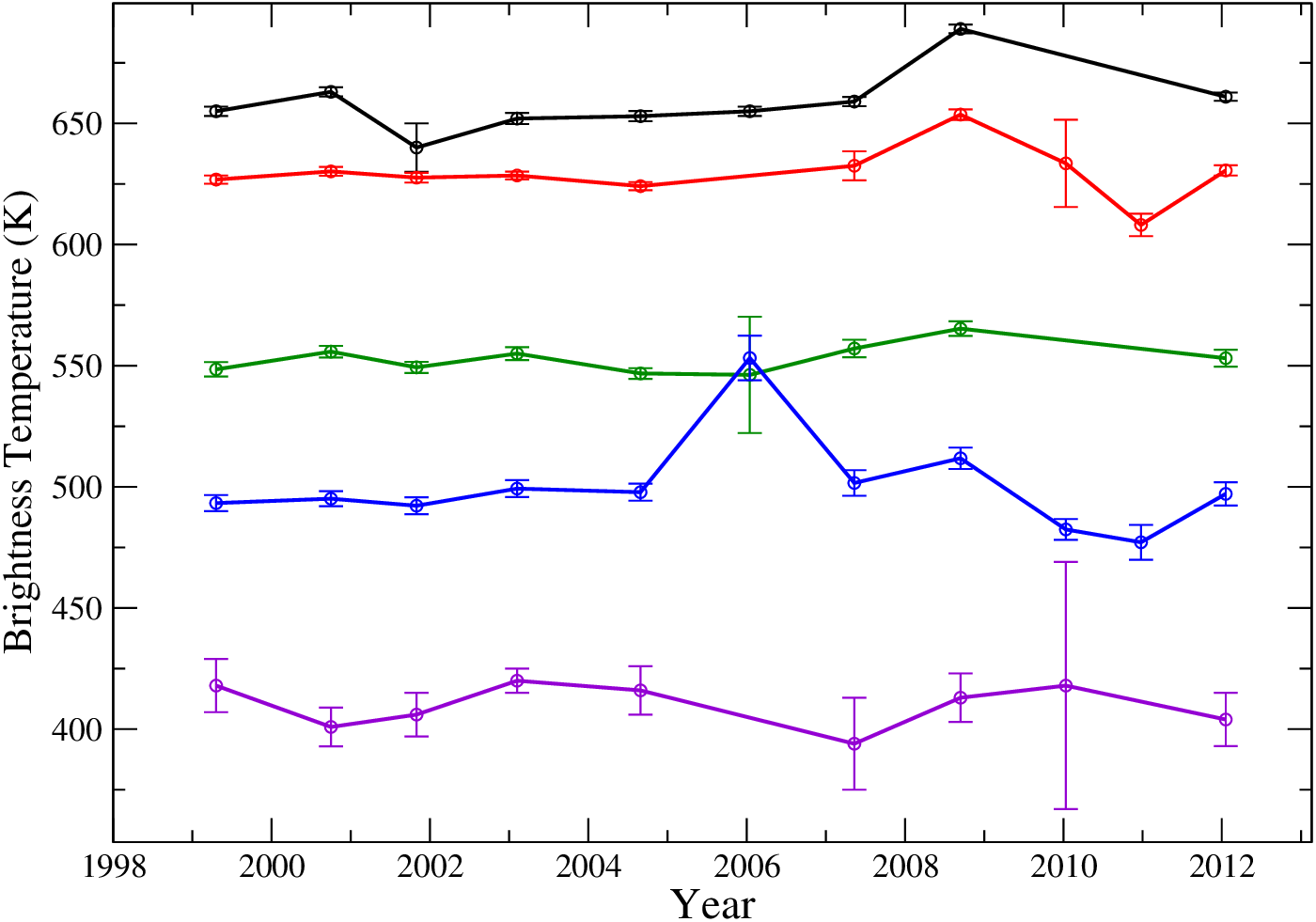}
\includegraphics[width=2in]{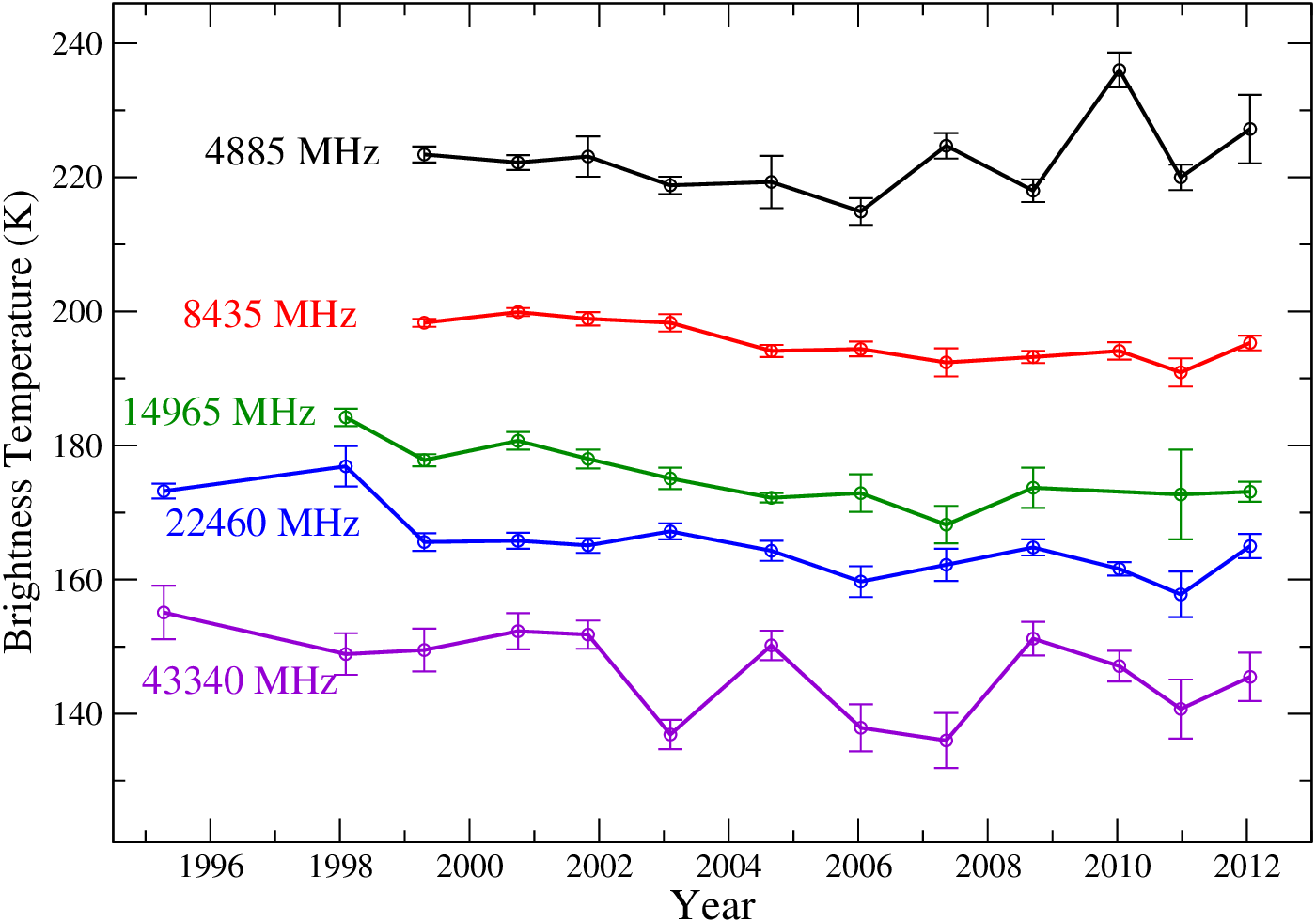}
\includegraphics[width=2in]{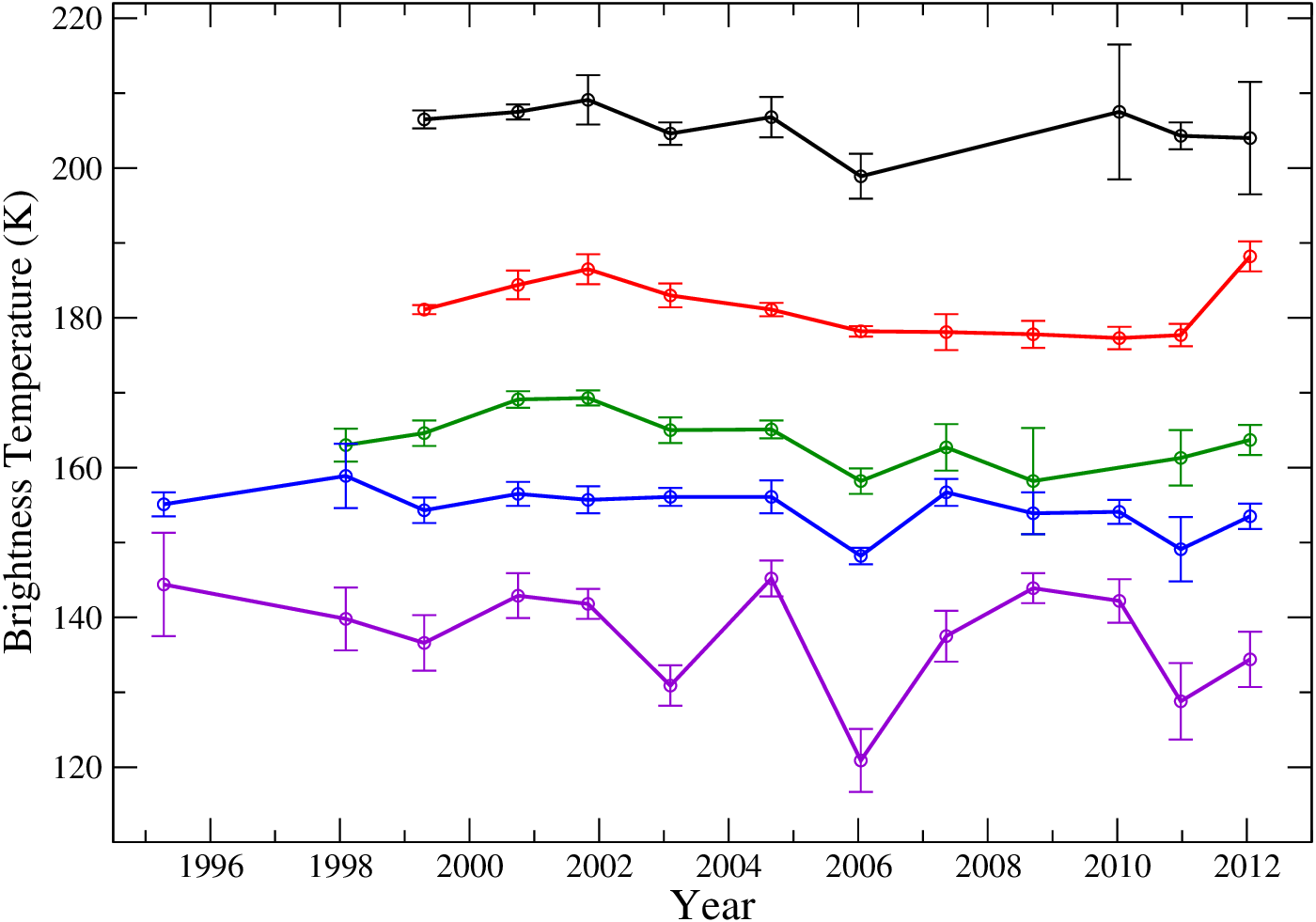}}}
\centerline{\parbox{6in}{
\caption{\small The secular variation in brightness for the three
  planets: Venus in the left panel, Uranus in the middle, Neptune on
  the right.  The small variation for Uranus is due to its polar
  regions coming more into view.}
\label{fig:TimePlanet}}}
\end{figure}

We added the planets Venus, Uranus, and Neptune (along with Mars) in
1995 in order to improve the model data for those objects.  We show in
Fig.~\ref{fig:PlanetSpectrum} their emission in terms of their brightness
temperature (properly accounting for the effect of the CMB background)
at selected frequencies, taken from the observations in January, 2012.  
The secular variations -- if any -- in the brightnesses at selected
frequencies are shown in Fig.~\ref{fig:TimePlanet}.

The emission from Venus is dominated at high frequencies by emissions
from the atmosphere and at low frequencies by emissions from the
surface; the crossover frequency is near 8 GHz, where the atmospheric
opacity is approximately 1 (\citet{Muh79}).  The flux density of Venus
at centimeter wavelengths has been described in \citet{But01}, and the
curious low-frequency turnover discussed in \citet{But04}.  Our
observations match the higher-frequency model calculations but the
decrease at low frequencies is still unexplained. Venus is a suitable
absolute calibrator for frequencies above 6 GHz, provided it is not
too large for the telescope or interferometer, and provided it is
not too near the Sun.

Radio emission from Uranus is entirely from the atmosphere, as the
sensible surface is too deep in the planet.  The flux
density of Uranus at centimeter wavelengths has been described in
\citet{Hof03} and at millimeter wavelengths by \citet{Gri93} (and
references therein for both).  Importantly, there is an enhancement of
emission from both poles, and so viewing geometry changes the flux
density significantly.  Given the $\sim$80 year orbital period of the
planet, along with its unusual pole position relative to the ecliptic,
variations on roughly decadal timescales are expected, and are clearly
detected (\citet{Hof03}; \citet{Kle06}).  The expected flux density
reached a minimum at the autumnal equinox in 2007, since the extent of
the polar enhancements was minimized then, and we see this reflected in
the measured flux densities.  There is still some uncertainty as to
whether these polar enhancements extend to millimeter wavelengths
(\citet{Hof07}).  Uranus is a suitable absolute calibrator at
higher frequencies (above a few GHz, depending on the sensitivity of
the telescope), up to the 50 GHz upper-limit of this study, as long as
these geometry considerations are taken into account

Neptune is similar to Uranus in that its emission is entirely from the
atmosphere.  The differences are in the lack of geometry changes for
Neptune (though there is also a south polar emission enhancement for
Neptune, its sub-Earth latitude does not change appreciably over even
decadal timescales), and differences in atmospheric composition
(\citet{DeB96}).  Neptune is a suitable absolute calibrator
at the frequencies of this study as long as the sensitivity of the
telescope is sufficient, though long-term studies of its emission are
not as complete as for Uranus.

Detailed comparisons of our observed brightness temperatures with
those predicted by various models are in preparation.  We expect that
these comparisons will result in improvements to the models of
characteristics of the surfaces and atmospheres of these planets which
will increase our understanding of them.

\section{A Comparision with the \citet{Baa77}, and other Scales}

One of the original purposes of this campaign was to determine the
accuracy of the \citet{Baa77} scale.  The results are given in
Table~\ref{tab:MarsBaars}, where we give the ratio at six frequencies
for the five sources 3C48, 3C123, 3C147, 3C286, and 3C295 between our
Mars-based fluxes and the corresponding \citet{Baa77} expressions.  As
three of the sources are shown here to be slightly variable, we have
for these objects utilized the average flux densities over the
measurement period.  Note that although the \citet{Baa77} scale for
these sources is not defined above 15 GHz, we have applied their
expressions to the higher frequency bands so users can conveniently
estimate any flux density scale errors introduced by the use of these
sources outside the recommended frequency range.  Unsurprisingly, the
deviations become quite large at these higher frequencies for most of
these sources.

\begin{deluxetable}{rcllllll}
\tabletypesize{\scriptsize} 
\tablecaption{Ratio of This Scale to the Baars Scale 
\label{tab:MarsBaars}}
\tablewidth{0pt} 
\tablehead{ \colhead{Source}& \colhead{327.5 MHz} &\colhead{1465 MHz}&
\colhead{4885 MHz}& \colhead{8435 MHz}& \colhead{14965 MHz}&
\colhead{22460 MHz}& \colhead{43340 MHz}} 
\startdata 
3C48 &1.02&1.037&1.006&1.002&1.034&1.11&1.24\\
3C123&1.12&1.007&0.946&0.918&0.899&0.89&0.86\\
3C147&1.06&1.000&0.944&0.949&1.000&1.10&1.44\\
3C286&1.00&1.021&0.987&0.976&0.978&0.99&1.04\\
3C295&1.04&1.035&0.979&0.972&1.000&1.05&1.21\\ 
\enddata 

\tablecomments{The ratios for the variable sources 3C48 and 3C147 are
  based on their long-term averages.  Note that our proposed scale, at
  327 MHz, is based on the spectral flux density of 3C196 given in
  \citet{Sca12}, and not on any absolute reference.} 
\end{deluxetable}

We also compared our new scale to the modified Baars scale proposed by
\citet{Ott94}, and to the 34 GHz values for 3C48, 3C147, and 3C286,
proposed by \citet{Mas99}, based on an absolute measurement made with
the 1.5m OVRO radio telescope. The ratio of our values to those of
\citet{Mas99} are 0.97, 1.03, and 0.94, for the three sources.  The
mean ratio of 0.98 is within the cited errors of $\sim$6\% for
\citet{Mas99} and $\sim$2\% for our scale.

\citet{Ott94} proposed modified versions of the Baars expressions for
3C286 and a number of other sources, using the Baars' values for 3C295
as a reference.  For our purpose, only their proposed expression for
3C286 is of interest.  We show in Table~\ref{tab:MarsBaarsOtt} the
values for selected frequencies.  Note that \citet{Ott94} utilized
various observations of 3C295 at high frequencies to extend the Baars
scale to 43 GHz. We conclude the high frequency extension of the Baars
scale by Ott is low by up to 10\% for frequencies above 15 GHz, with
the error increasing with increasing frequency.

\begin{deluxetable}{ccccccc}
\tabletypesize{\scriptsize}
\tablecaption{A comparison of ours, Baars, and Ott scales for 3C286
\label{tab:MarsBaarsOtt}}
\tablewidth{0pt}
\tablehead{\colhead{Scale}&\colhead{1465 MHz}&
\colhead{4885 MHz}& \colhead{8435 MHz}& \colhead{14965 MHz}&
\colhead{22460 MHz}& \colhead{43340 MHz}} 
\startdata 
Baars&14.51&7.41&5.19&3.45&2.53&1.47\\
Ott  &14.36&7.47&5.18&3.38&2.42&1.35\\
This paper &14.81&7.31&5.07&3.37&2.51&1.53\\
\enddata
\end{deluxetable}

\section{Summary}

The VLA, when used with care, is capable of measuring the flux density
ratios between compact, bright, and isolated radio sources with an
accuracy much better than 1\% at most frequency bands. We have
utilized this capability to measure the ratios between a set of
proposed calibration sources, covering the entire frequency range from
1 through 50 GHz. The observations span more than 30 years at some
frequencies.  The set of observed sources included seven compact
extragalactic sources, two galactic planetary nebulae, one evolved
star, and four planets.

The VLA cannot make accurate absolute measurements of the spectral
flux density of radio sources.  We converted our accurate ratio measurements to
spectral flux densities by utilizing a thermophysical emission model
of the planet Mars.  The model was placed on an absolute
scale by utilizing the WMAP observations of Mars, which are calibrated
on the CMB dipole anisotropy.

>From the nine compact, non-planetary objects, we determine that four
sources -- 3C123, 3C196, 3C286, and 3C295 are stable to within 1\%
over the 30-year span of this program at all frequencies except above
$\sim$ 40 GHz, where our accuracy is limited by antenna pointing.  We
present polynomial expressions for the spectral flux density of these
four sources.  Of these, 3C286 is the most compact with the
flattest spectrum, on which basis we have selected it as our
interferometric standard flux density calibrator.  Using its derived
spectrum, we determine the spectral flux densities of all the other
sources over the frequency range of 1 to 50 GHz.

The three sources 3C48, 3C138, and 3C147 are commonly used flux
density calibrators for the VLA and other interferometers.  However,
all are variable on timescales of several years.  We have fitted
polynomial expressions to their spectral flux densities as a function
of time, to allow past users of the VLA to correct their calibration.  

The planets Venus, Uranus, and Neptune are all potential calibration
sources at high radio frequencies.  Using our new scale, we have
derived their mean brightness temperatures, to assist in improving
models of their atmospheric emission.  

The absolute accuracy of the new scale is estimated at $\sim$2\%, with
the largest source of error being the transfer of the WMAP
observations to the VLA radio observations.  The errors in the
internal transfer from the derived spectral flux density of 3C286 to
other sources is estimated at less than $\sim$1\% for all frequencies
below $\sim$20 GHz, and rises to $\sim$3\% at the highest VLA
frequency of 50 GHz.  

We emphasize that this very low error does not mean that all
observations with the VLA have this accuracy.  The actual accuracy
obtained for any observation will, in addition to the estimated
accuracy of the flux density scale, be determined by how the
observation is set up, the observing conditions, and the care taken in
the calibration and imaging steps.

The authors thank Eric Greisen for the many improvements made to the
\texttt{AIPS} data reduction software package over the years of the
project, particularly those needed in response to the dramatic
increase in the VLA's capability with the implementation of the WIDAR
correlator.

\end{document}